\documentclass[
showpacs,preprintnumbers,amsmath,amssymb,prb]{revtex4}
\usepackage{graphicx}
\usepackage{dcolumn}
\usepackage{bm}
\usepackage{amssymb}
\usepackage{amsmath}
\usepackage{epsf}

\begin{document}

\title{Non-Markovian Effects on the Two-Dimensional Magnetotransport: Low-field Anomaly in
Magnetoresistance }
\author{Vadim~V.~Cheianov$^{1}$, A.~P.~Dmitriev$^{2}$, V.~Yu.~Kachorovskii$^{2}$}
\affiliation{$^1$NORDITA, Blegdamsvej 17, Copenhagen,  DK 2100, Denmark
\\ $^2$A.F.~Ioffe Physical-Technical Institute, 26
Polytechnicheskaya str., Saint Petersburg, 194021,
Russia}

\date{\today}

\begin{abstract}

 We discuss  classical magnetotransport in a two-dimensional
 system with strong scatterers. Even in the limit of very low field, when
 $\omega_c \tau \ll 1$ ($\omega_c$ is the cyclotron frequency,
 $\tau$ is the scattering time) such a system demonstrates
 strong negative magnetoresistance caused by non-Markovian memory
 effects. A regular method for the calculation of non-Markovian
 corrections to the Drude conductivity is presented. A quantitative
 theory of the recently discovered anomalous low-field
 magnetoresistance is developed for the system of two-dimensional
 electrons scattered by hard disks of radius $a,$ randomly
 distributed with concentration $n.$ For small magnetic fields the
 magentoresistance is found to be parabolic and inversely
 proportional to the gas parameter, $ \delta \rho_{xx}/\rho \sim -
 (\omega_c \tau)^2 / n a^2.$ In some interval of magnetic fields
 the magnetoresistance is shown to be linear $\delta
 \rho_{xx}/\rho \sim - \omega_c \tau $ in a good agreement with the
 experiment and numerical simulations. Magnetoresistance saturates
  for $\omega_c \tau \gg na^2$, when  the anomalous memory effects
 are  totally destroyed by the magnetic field. We also discuss
 magnetotransport at very low fields and
 show that at such fields magnetoresistance is determined by the
 trajectories having a long Lyapunov region.
\end{abstract}


\maketitle
\section{Intorduction}
\narrowtext

      The problem of magnetoresistance in metal and semiconductor
structures has been intensively discussed in literature during the
past three decades. A large number of both theoretical and
experimental papers on this subject was published. Most of these
works were devoted to the case of the degenerate two dimensional
electron gas where the electrons move in the plane perpendicular
to the magnetic field and scatter on a random impurity potential.
In this situation only the electrons with energy close to the
Fermi energy participate in conductance and the usual approach
based on the Boltzmann equation leads to vanishing
magnetoresistance. In other words, the longitudinal resistance
$\rho_{xx}$ of the sample does not depend on the magnetic field
$B.$ This implies that the explanation of the experimentally
observed $B$ dependence of the longitudinal resistance should be
sought beyond the Boltzmann theory.

The intense exploration of this area began from the work of Altshuler {\it et al}, \cite{Alt} where
the experimentally observed in 2D metals and semiconductor structures negative magnetoresistance
(MR), i.e. decreasing $\rho_{xx}$ with increasing $B$, was explained by quantum interference
effects. It was shown that the magnetic field destroys the negative weak localization correction to
the conductivity, thus resulting in decreasing longitudinal resistance. Since the first publication
on the subject \cite{Alt} a vast amount of work has been devoted to its further exploration (see
for review Ref.~\onlinecite{lee}).

Two years prior to Ref.~\onlinecite{Alt} there appeared a publication \cite{baskin} where a
classical mechanism of negative magnetoresistance was discussed. The mechanism was investigated on
the example of a gas of non-interacting electrons scattering on hard disks (antidots). It was shown
that with increasing magnetic field there is an increasing number of closed electron orbits which
avoid scatterers and therefore are not diffusive (see also recent discussions
  \cite{boby,bobynew,basknew} of this mechanism). Electrons occupying these orbits do not participate
in diffusion. As a result, the longitudinal resistance turns out to be proportional to the factor
$1-P,$ where $P=\exp(-2 \pi /\omega_c \tau)$ is the probability of the existance of the circular
closed orbit, which avoids scatterers (here $\omega_c$ is the cyclotron frequency, $\tau$ is the
scattering time). Another classical mechanism was presented in Ref.~\onlinecite{polyakov}, where
the MR due to non-markovian dynamics of electrons  trapped in some region of space was discussed.

Notwithstanding these developments, the role of classical effects in magnetotransport was
underappreciated for a long time. A new boost to the research in this direction was given by
Ref.~\onlinecite{perel}, where it was shown that if electrons move in a smooth disorder potential
and in a sufficiently strong magnetic fields a phenomenon called "classical localization" occurs.
This phenomenon leads to the exponential suppression of the longitudinal resistance: most electrons
are trapped in localized equipotential trajectories and do not participate in diffusion. This work
was followed by a series of works, \cite{fog,mir1,mir2,mir3,igor,dmit,dmit1,my} discussing
different aspects of classical magnetotransport in 2D systems.
   It was
shown \cite{mir1} that for lower magnetic fields near the onset of the classical localization the
magnetoresistance is positive, i.e. the longitudinal resistance grows with increasing magnetic
field. In Refs.~\onlinecite{mir2,mir3} the combination of smooth disorder and strong scatterers
(antidots) was considered. It was shown that in this system under certain conditions there are
several regimes of the behavior of magnetoresistance  depending on the strength of the magnetic
field: first the longitudinal resistance decreases with growing field, then it saturates and then
begins to grow.

In Refs.~\onlinecite{baskin,boby,bobynew,basknew,kuzm, polyakov,perel,fog,mir1,mir2,mir3,igor}
magnetoresistance was studied in a situation where the magnetic field is classically strong, that
is where the parameter $\beta=\omega_c\tau$ is large. Recently,  the region of classically small
magnetic fields $\beta\ll 1 $ was investigated numerically \cite{dmit,dmit1} for the case of
electrons scattering on strong scatterers. It was shown \cite{dmit} that memory effects due to
double scattering of an electron on the same disk lead to a negative parabolic magnetoresistance
(in the Ref.~\onlinecite{baskin}, where these processes were not taken into account, exponentially
small MR was predicted).
The numerical simulations \cite{dmit1} discovered a low-field classical anomaly of the MR. The
anomaly was attributed to the memory effects specific for backscattering events.  The simulations
were performed for the 2D Lorenz gas which is a system of 2D electrons scattering on hard disks
randomly distributed in plane with average concentration $n.$ Magnetotransport in this system is
characterized by two dimensionless parameters: $ \beta= \omega_c \tau ,$ and the gas parameter
$\beta_0=a/l=2 n a^2 .$ Here $a$ is the disk radius, $\omega_c$ is the cyclotron frequency,
$\tau=l/v_F$ is the mean free time and $l=1/2 n a$ is the mean free path. The anomaly was observed
in the case $\beta \ll 1 , \quad \beta_0 \ll 1.$ Both the numerical simulations and the qualitative
considerations  \cite{dmit1} indicated that at zero temperature the MR can be expressed in terms of
a dimensionless function $f(z)$ via
\begin{equation} \frac{\delta\rho_{xx}}{\rho}
=-\beta_0 f\left(\frac{\beta}{\beta_0}\right),
\label{1}
\end{equation}
where $\rho $ is the resistivity for $B=0.$ Numerical results
\cite{dmit1} suggest that $f(z)\sim z$ as $z\to 0,$ yielding
\begin{equation}
\frac{\delta\rho_{xx}}{\rho} \sim -|\omega_c|\tau .
\label{lin}
\end{equation}
The latter expression is in a very good agreement with experimental measurements of negative linear
MR in a random antidot arrays.\cite{dot1}
It is anomalous in two senses. First, it
has a non-analytic dependence on the magnetic field. Second, it
does not vanish in the limit of vanishing $\beta_0$, which is
normally regarded as the expansion parameter for the corrections
to the Drude-Boltzmann picture. This intriguing behavior calls for
a rigorous analytical theory of the effect, which would establish
Eq.~\eqref{1} and enable one to derive the analytical expression
for function $f$.

In this paper we present a detailed theory of the anomaly and give an expression for $f(z)$ (the
brief description of our results was given  in Ref.~\onlinecite{my}).
 We find that at some interval, $0.05 \lesssim z \lesssim 2,$
function $f(z)$ is linear in agreement with numerical simulations
and experiment, but at  $z \to 0$ crosses over to a quadratic
dependence. Thus, for $\beta\to 0,$ Eq.~\eqref{1} yields $\delta
\rho_{xx}/\rho \sim -\beta^2/\beta_0.$ The limit $\beta_0\to 0$
should be taken with care. While the small $\beta$ expansion seems
to be singular as a function of $\beta_0,$ the region of $\beta$
where this expansion is valid shrinks as $\beta_0\to 0.$ For $z
\to \infty,$ $f$ saturates at some constant value. Therefore, the
full variation of $\delta \rho_{xx}/\rho$ is of the order
$\beta_0.$ In other words, the anomalous MR is strong but it
exists in a small region of magnetic fields.

\section{Qualitative discussion of the problem \label{Qualitative} }
\begin{figure}
\includegraphics[width=0.5\textwidth]{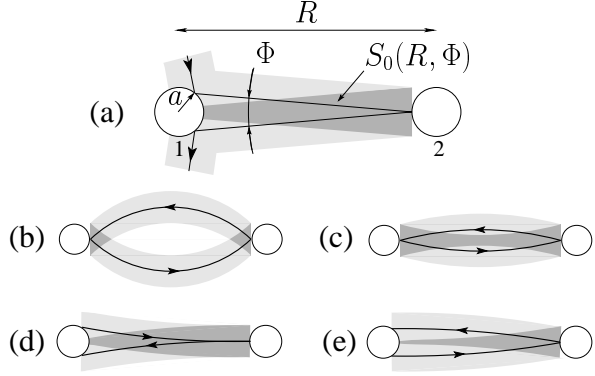}
\caption{ Backscattering process responsible for leading
nonanalytic contribution to the resistivity at $B=0$ (a).  For $B
\neq 0,$ the overlap area, $S_B,$ between two corridors is small
at large $B$ (b). For $\Phi=0,$ $S_B$ decreases with $B$ (c). For
$\Phi \neq 0$ and small $B$ the values of $S_B- S_0$ for time
reversed trajectories have opposite signs (d,e).} \label{fig1}
\end{figure}

The mechanism proposed in Ref.~\onlinecite{dmit1} is linked to the memory effects arising in
backscattering events. It has a close relation to the well known non-analyticity of the virial
expansion of transport coefficients, \cite{dorfman,wei,wei1,hauge,peierls} which we briefly recall.
For $B=0$ the leading nonanalytic correction to resistivity, $\delta \rho$, is due to the processes
of return to a scatterer after a single collision on another scatterer (see Fig.~\ref{fig1}a).
The relative correction, $\delta \rho/\rho,$ is proportional to
the corresponding backscattering probability, given by the product
of $ e^ {-R/l} d \Phi dR/l $ (which is the probability to reach
scatterer 2 without collision and scatter in the angle $d\Phi$)
and the probability $p$ to return without collisions from 2 to 1
(here $l$ is the mean free pass).  Assuming $p=\exp(-R/l)$ and
integrating over intervals $ 0< \Phi< a/R, \quad a<R<\infty,$ one
obtains \cite{dorfman,wei,wei1, hauge,peierls}
\begin{equation}
\delta \rho /\rho\sim
\int_a^{\infty} \frac{dR}{l} \int_0^{a/R} d\Phi \ e^{-2R/l}
\sim \beta_0 \ln(1/2\beta_0).
\label{Weil}
\end{equation}

In Ref.~\onlinecite{dmit1} it was shown that the probability $p$ is actually larger than
$\exp(-R/l).$ Indeed, the exponent $\exp(-R/l) $ can be written as $\exp(-nS),$ where $S= 2a R.$ It
represents the probability of the existence of an empty corridor (free of the centers of the disk)
of width $2a$ around the electron trajectory from $2$ to $1.$ However, the passage of a particle
from $1$ to $2$ ensures the absence of the disks centers in the region of width $2a$ around this
part of trajectory (from $1$ to $2$) This reduces the scattering probability on the way back. The
correct value of $p$ can be estimated as
\begin{equation}
    p(R,\Phi)=\exp(-n(S-S_0)) = \exp(-R/l +n S_0(R,\Phi)),
    \label{p}
\end{equation} where
\begin{equation}
 S_0(R, \Phi) = 2 a R -R^2
|\Phi|/2 \label{s00}
\end{equation}
 is the area of the overlap of the two corridors
(see Fig.~\ref{fig1}a).  For example, for $\Phi=0,$ we have
$S_0=2aR$ and $p=1,$ which reflects the obvious fact that
the particle cannot scatter, if it travels back along the same path.
Taking into account the effect of ``empty corridor'', we get
\begin{equation} \frac{\delta \rho}{\rho} \sim
\int_a^{\infty} \frac{dR}{l} \int_0^{a/R} d\Phi \ e^{-(2R/l)
+nS_0} \approx \beta_0 \ln \left(\frac{C} {2 \beta_0}\right),
\label{estimate} \end{equation}
where $C$ is a constant of the order of unity.  Thus,
for $B=0$ the "empty corridor" effect simply changes the constant in
the argument of the logarithm.

The key idea suggested in Ref.~\onlinecite{dmit1} was that for $B\neq 0$ the area of the overlap of
the two corridors, $S_B,$ sharply depends on $B,$ resulting in the observed MR.  Indeed, it is seen
from Fig.~\ref{fig1}b that for $\beta \gtrsim \beta_0 ,$ $S_B \to 0,$ resulting in sharp negative
MR
\begin{equation} \frac{\delta \rho_{xx}}{\rho} \sim
\int_0^{\infty} \frac{dR}{l} \int_0^{a/R} d\phi \ e^{-2R/l} \left
( e^{n S_B} -e^{n S_0} \right).  \label{estimateB} \end{equation}
The following qualitative explanation of the observed linear MR was presented in
Ref.~\onlinecite{dmit1}. The value $n ( S_B - S_0)$ was estimated for $\phi=0$ (see
Fig.~\ref{fig1}c) to the first order in $B$ as $- nR^3/R_c =- R^3/2 a l R_c,$ where $R_c$ is the
cyclotron radius. Assuming that this estimate also works at $\phi \neq 0$ and expanding $e^{n S_B}
- e^{n S_0}$ to the first order in $B,$ one gets $ \delta \rho_{xx}/ \rho \sim -l/R_c = - \omega_c
\tau.$

In fact, the physical picture of the phenomenon is more subtle. The contribution of any trajectory
with $\phi \neq 0$ is cancelled to the first order in $B $ by the contribution of the time-reversed
trajectory, since the values of $ S_B - S_0$ are opposite for these paths (see Fig.~1~d,~e).  The
cancellation does not occur only at very small $\phi \sim \beta.$ The integration in
Eq.~\eqref{estimateB} over $\phi < \beta$  yields $\delta \rho_{xx} /\rho \sim - \beta^2/\beta_0 $.
Larger values of $\phi$ also give a quadratic in $\beta$ contribution to the MR. This contribution
is positive and comes from the second order term in the expansion of $e^ {n S_B}- e^{nS_0}$ in $B$.
It follows from our results (see Eqs.~\eqref{1},\eqref{f(z)}) that the contribution of small angles
is dominant resulting in a negative parabolic MR.
 We find that the parabolic MR crosses
over to linear at very small $\beta \approx 0.05 \beta_0, $ which explains why the parabolic MR was
not seen in numerical simulations \cite{dmit1} and experiment. \cite{dot1} Note that for very small
$\beta<\beta_0^2$ contribution of the trajectories with a long Lyaponov region (Lyapunov
trajectories) becomes important. In the Section~\ref{calc} we focus on the region $\beta>\beta_0^2$
where the contribution of the Lyapunov trajectories is parametrically small. The Lyapunov
trajectories will be discussed   in Section~\ref{lyap}.

\section{Calculations \label{calc}}

\subsection{Kinetic equation}
In this subsection we introduce the kinetic equation which
is the starting point for the calculation of the diffusion
coefficient and the resistivity of the Lorenz gas.

We consider the Lorenz gas at zero temperature, assuming that
the electrons participating in the conduction have the Fermi
velocity $v_F.$ The diffusion coefficient $D$ is given by
\begin{equation}
D=\frac{1}{2}\int_0^\infty d t \langle \mathbf v(0)
\mathbf
v(t)\rangle= \frac{1}{2}\int_{-\infty}^{\infty}dt
\int d\mathbf{r} d\mathbf{v}
\langle G_R \rangle
\mathbf v \mathbf v_0.  \label{D} \end{equation}
Here $G_R=G_R(\mathbf v, \mathbf v_0,\mathbf{r}, \mathbf r_0,t)$ is the
 retarded
Green's function of the Liouville equation and
$\langle\dots\rangle$ stands for the averaging over
the positions of the disks.  The equation for $G_R$
reads
\begin{align}
& \left(\frac{\partial }{\partial t} +  \mathbf v
\frac {\partial }{\partial \mathbf r}
 - \omega_c [\mathbf v \times \frac{ \partial } {\partial
\mathbf v}] - \hat T
\right) G_R =
\delta(\mathbf r -\mathbf r_0) \delta(\mathbf v - \mathbf v_0 ) \delta(t),
\nonumber
\\
& G_R=0, \qquad \mathrm{for} \quad t<0, \label{LL}
\end{align}
where $ \mathbf{v} \partial /\partial
\mathbf{r}
- \omega_c [\mathbf v \times \partial /\partial
\mathbf v]$
is the Liouville operator of the free motion in the
magnetic field and $\hat T$ describes scattering on the
disks. From Eq.~\eqref{D} it follows that we need the time
integral of $G_R$ rather than the whole time-dependent function.
This integral can be written as
\begin{equation}
\int_{0}^{\infty} dt G_R(\mathbf v, \mathbf v_0,\mathbf
r,\mathbf r_0,t)=\frac{\delta(v-v_F)}{v_F} ~ G_{\alpha,
\alpha_0}( \mathbf r, \mathbf r_0 ),
\label{intGt}
\end{equation}
where $\alpha, \alpha_0$ are the angles of velocities
$\mathbf v$ and $\mathbf v_0$ respectively. Here we used the
energy conservation, which implies that in scattering processes
the absolute value of the velocity does not change. Using Eq.~\eqref{intGt},
the diffusion coefficient Eq.~\eqref{D} can be rewritten as
\begin{equation}
D=\frac{v_F}{4\pi \rm{ S}} \int d \mathbf r d\mathbf r_0 d \alpha d \alpha_0
\langle G_{\alpha, \alpha_0}(
\mathbf r, \mathbf r_0 )  \rangle \cos (\alpha-\alpha_0).
\label{D1}
\end{equation}
Since $\langle G_{\alpha, \alpha_0}(
\mathbf r ,\mathbf r_0)  \rangle$ depends on $\alpha-\alpha_0$ and $\mathbf r -\mathbf
r_0$ only,
it is convenient to average over the
position of the initial point and over the initial angle. Here
${\rm
S}$ is the  area of the sample.
Integrating Eq.~\eqref{LL} with respect to time
the equation for $G=G_{\alpha, \alpha_0}(
\mathbf r ,\mathbf r_0)$ is written as
\begin{equation}
\left(\epsilon + \hat {
 L}_0- \hat
T_{-}-\hat T_{+} \right) G =
\delta(\mathbf r - \mathbf r_0)\delta(\alpha -\alpha_0),
\label{Lio}
\end{equation}
where $$ \hat L_0=  \mathbf{n} \frac{\partial}{\partial \mathbf r} - \frac{1}{R_c}
\frac{\partial}{\partial \alpha}, $$ $\mathbf n=\cos\alpha \mathbf e_x +\sin\alpha \mathbf e_y$ is
the unit vector in the direction of $\mathbf v$ and $\epsilon \to 0.$   The interaction with disks
is written in Eq.~\eqref{Lio} in the form of collision integral. \cite{wei,mori}  The scattering
operators $\hat T^{\pm}$ transform arbitrary function $f(\mathbf r, \alpha)$ as follows,
\begin{align}  &\hat T^+ f(\mathbf r,\alpha)=
\int d \alpha^{\prime} \sigma(\alpha - \alpha^{\prime})
n^+ f(\mathbf r, \alpha^{\prime}),
\nonumber
\\
\label{+,-}
&\hat T^- f(\mathbf r ,\alpha)= - f(\mathbf{r}, \alpha ) \int d \alpha^{\prime}
\sigma(\alpha - \alpha^{\prime})
n^- ,
\end{align}
where
\begin{equation}
n^{\pm} =n^{\pm}(\mathbf r, \mathbf a) =\sum_i \delta(\mathbf{r}-
\mathbf R_i \pm \mathbf{a} ). \label{npm}
\end{equation}
\begin{figure}
\includegraphics[width=0.25\textwidth]{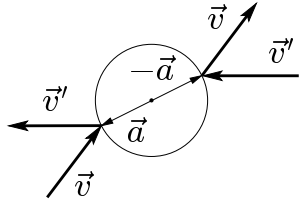}
\caption{Scattering of a particle on a hard disk. }
\label{fig2}
\end{figure}
Here $\mathbf R_i$ are the positions of the disk's centers and
\begin{equation}
\sigma(\varphi) = \frac{a}2 \left\vert\sin
\frac{\varphi}2\right\vert
\label{crossection}
\end{equation}
is the differential cross-section of one disk. The
vector $\mathbf a$ depends on angles $\alpha,~\alpha'$ in the
integrals Eq.~\eqref{+,-}
\begin{equation}
\mathbf a= \mathbf a_{\alpha, \alpha' } =\frac{ (\mathbf
n^{\prime} - \mathbf n)a}{\sqrt{2(1- \mathbf n^{\prime} \mathbf
n)}} , \label{adef}
\end{equation}
and is pointing from the center of a disk to the scattering point
at the disk surface (see Fig.~\ref{fig2}). Physically, operator
$\hat T^{+}$ describes influx of particles to velocity ${\bf v}$
at the point ${\bf R}_i - \bf{a}$, while operator $\hat T^{-}$
describes the outflux from velocity $ \bf{ v} $ at the point
$\mathbf R_i + \mathbf{a}.$ The averaged value of the $n^{\pm}$ is
equal to the disks concentartion
\begin{equation}
\langle n^{+}\rangle = \langle n^{-}\rangle =
n.
\label{naverage}
\end{equation}

The Bolzmann equation is obtained by averaging the Liouville
equation with respect to the position of the scatterers and neglecting
correlations. Indeed, using Eq.~\eqref{naverage} one finds that
the average collision operators Eq.~\eqref{+,-} are given by
\begin{align}\nonumber
& \langle \hat T^+\rangle f(\mathbf r ,\alpha)= n \int d
\alpha^{\prime} \sigma(\alpha -\alpha^{\prime}) f(\mathbf r,
\alpha^{\prime} ),
\\
\label{tt} & \langle \hat T^-\rangle =-n\int d \alpha' \sigma(\alpha-\alpha')= - 1/l.
\end{align}
Replacing the collision operators in the Liouville equation Eq.~\eqref{Lio}
with their averages Eq.~\eqref{tt} one obtains the Boltzmann equation
\begin{equation}
 \hat  L_0  G^B_{\alpha, \alpha_0}(\mathbf r) -
 n\int  d \alpha'
\sigma (\alpha - \alpha')
(G^B_{\alpha', \alpha_0}(\mathbf r) -
G^B_{\alpha, \alpha_0}(\mathbf r)) =
\delta(\mathbf{r})\delta(\alpha -\alpha_0). \label{bol}
\end{equation}
The Bolzmann-Drude diffusion coefficient $D_0$ in the
absence of the magnetic field is
found from Eqs. \eqref{bol} and \eqref{D1} as follows.
Integrating \eqref{bol} with respect to $\mathbf r$ and
expanding the Green's function $G^B$ in angular harmonics one
finds
\begin{equation}
\int G^B_{\alpha,\alpha_0}(\mathbf r) d \mathbf r = \frac{1}{2\pi}
\sum_m l_m e^{i m (\alpha - \alpha_0)}, \label{intg}
\end{equation}  where
\begin{equation}
l_m^{-1}= n \int d \varphi \sigma(\varphi)\left(1-\cos (m
\varphi)\right), \label{lm}
\end{equation}
In particular,
\begin{equation}
l_1=l_{\rm tr}=\frac{3 l}{4}=\frac{3}{8na}
\label{ltrdef}
\end{equation}
is the transport length.
Substituting Eqs.~\eqref{intg} and \eqref{ltrdef} in Eq.~\eqref{D1}
one finds
\begin{equation}
D_0=\frac{v_F l_\text{tr}}{2}
\end{equation}
\subsection{Perturbative expansion}
In this subsection we derive the perturbative expansion for the
average Green's function in order to take into account non-Markovian
corrections which are absent in the Bolzmann-Drude picture.

Introduce the operators
\begin{equation}
\delta \hat T^{\pm} = \hat T^{\pm} - \langle \hat
T^{\pm} \rangle
\label{Tfluke}
\end{equation}
which describe the fluctuation of the collision integral
with respect to its average value.
Using these operators the  formal solution of \eqref{Lio},
$\hat G =(\epsilon +\hat L_0 -\hat T^- -\hat T^+)^{-1}$ can be
written as the following series
\begin{equation} \langle \hat G \rangle= \hat G^B
+\sum_{\mu,\nu = \pm} \hat G^B \langle \delta
\hat T^{\mu}\hat  G^B
\delta \hat T^{\nu} \rangle \hat G^B + \cdots,
\label{series}
\end{equation}
where $G^B$ is defined by Eq.~\eqref{bol}.
Here we took into account that $\langle \delta T \rangle=0.$
The first term on the right hand side of Eq.~\eqref{series} gives
the Drude-Boltzmann result described in the previous subsection.
The rest of the terms in Eq.~\eqref{series} provide a regular way for
the calculation of correlations, which are absent in the
Boltzmann picture.

\begin{figure}
\includegraphics[width=0.55\textwidth]{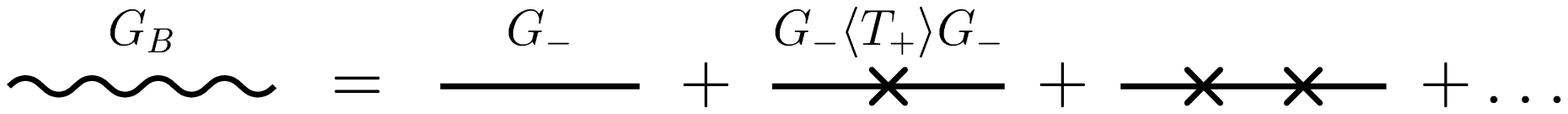}
\caption{Expansion of Boltzmann propagator $G^B$ in a series in
ballistic propagators $G^-$ } \label{fig4}
\end{figure}

In the subsequent analysis of the perturbative expansion
Eq.~\eqref{series} we will extensively use the representation of
the Boltzmann propagator $G_B$ in terms of the ballistic propagator
\begin{equation}
\label{bal}
\hat G^- = ( \hat L_0 + 1/l)^{-1}.
\end{equation}
This is achieved by expanding the Boltzmann
propagator $G^B=(L_0+1/l-\langle T^{+}\rangle)^{-1}$ as
a sum over the number of scattering events represented by the operator
$\langle T^{+}\rangle$
\begin{equation}
G^B=G^{-}+G^{-}\langle T^{+}\rangle G^{-}+\dots.
\label{GBexp}
\end{equation}
The diagrammatic representation of the expansion \eqref{GBexp} is shown in Fig.~(\ref{fig4}).
 Each cross in this figure corresponds to $\langle
\hat T^+\rangle$ and each solid line   to a ballistic propagator
$G^{-}.$ For future reference we write the explicit expression for
the ballistic Green's function in zero magnetic field and in weak
magnetic fields. For $B=0$ the ballistic propagator conserves the
velocity. In this case its kernel is given by
\begin{equation}
G^-_{ \alpha,  \alpha_0}( \mathbf r) = \delta( \alpha - \alpha_0)
\delta(\alpha_0 - \alpha[\mathbf r])\frac{ \exp\left(-r/l\right)
}{r} \label{G-}.
\end{equation}
Here $\alpha[\mathbf r]$ is the angle of the vector $\mathbf r$.
Magnetic field rotates the velocity vector with the cyclotron  frequency.
For small magnetic fields, $\beta \ll 1$, we have
\begin{equation}
G^-_{ \alpha,  \alpha_0}( \mathbf r) \approx \delta( \alpha -
\alpha[\mathbf r]-r/2R_c)
\delta(\alpha_0 - \alpha[\mathbf r]+r/2R_c)\frac{
\exp\left(-r/l\right) }{r} \label{G-B}.
\end{equation}

\subsection{Ballistic returns in perturbative expansion}
\begin{figure}
\includegraphics[width=0.5\textwidth]{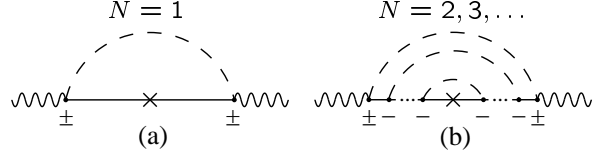}
\caption{Diagrams, corresponding to the process shown in Fig.~\ref{fig1}a.  Diagram (a) does not
take into account the ``empty corridor'' effect and should be renormalized by diagrams (b). }
\label{fig3} \end{figure}

 In this subsection we show how the processes of ballistic returns
discussed in Section \ref{Qualitative} arise in the perturbative expansion Eq.~\eqref{GBexp}.
  Consider the second term in Eq.~\eqref{series}. This term
describes the memory effect due to diffusive returns. As discussed in Section \ref{Qualitative},
the main contribution comes from returns after a single scattering. This process is described by
the diagram  shown in Fig.~\ref{fig3}a. The dashed line corresponds to the pairings $\langle \delta
\hat T^{\mu} \delta \hat T^{\nu}\rangle $ ($\mu,\nu =\pm$), external wavy lines to the diffusion
propagators $\hat G^B.$ The internal line corresponds to the Boltzmann propagator Eq.~\eqref{GBexp}
truncated at one scattering $\hat G^- \langle \hat T^+ \rangle \hat G^-.$
\begin{figure}
\includegraphics[width=0.35\textwidth]{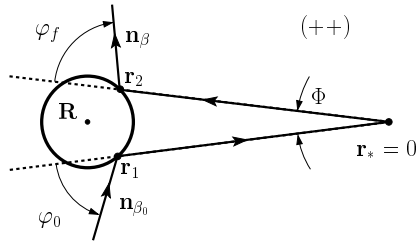}
\caption{Electron trajectory related to the pairing $(+,+)$. An electron   scatters twice (at
points $\mathbf r_1$ and $\mathbf r_2$)  on the surface of a disk, placed at point $\mathbf R.$ For
a fixed position of the backscattering point ($\mathbf r_*=0$), the process is parameterized by
$\mathbf R$, and by the scattering angles $\varphi_0,\varphi_f$} \label{fig6}
\end{figure}
\begin{figure}
\includegraphics[width=0.35\textwidth]{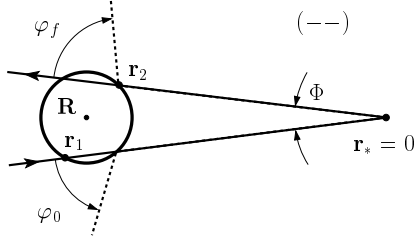}
\caption{Electron trajectory for the $(-,-)$ pairing. Electron does not experience any real
scattering except as scattering at $\mathbf r_*=0$. The diagram $(-,-)$ gives the probability to
pass twice the region around point $\mathbf R$ without scattering.  } \label{fig9}
\end{figure}
\begin{figure}
\includegraphics[width=0.35\textwidth]{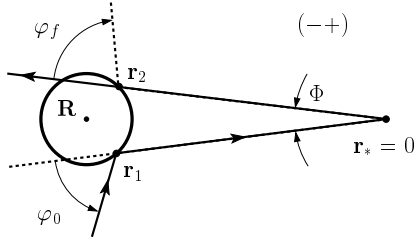}
\caption{ "Unphysical" electron trajectory related to the pairing $(-,+)$: electron  scatters on a
disk and after a while passes the region occupied by this disk without scattering.} \label{fig8}
\end{figure}
\begin{figure}
\includegraphics[width=0.35\textwidth]{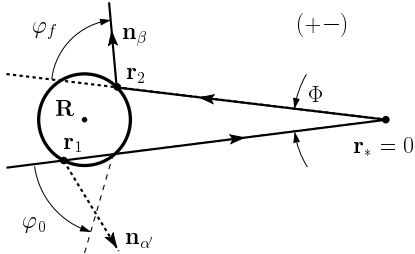}
\caption{"Unphysical" electron trajectory related to the pairing $(+,-)$: electron passes the
region occupied by a disk without scattering  and later on scatters on this disk. The diagram
related to such trajectory corrects the Boltzmann result by subtracting the probability of this
unphysical event. It is convinient to parameterize this trajectory by
 the scattering angles for the $(+,+)$
process.} \label{fig7}
\end{figure}
Four combinations of $(\pm,\pm)$ at the ends of external dashed lines in the diagrams shown in
Fig.\ref{fig3}a represent
 four different types of correlation at a given point $\mathbf R.$ To see this consider the pairing
$\langle \delta \hat T^{\mu} \delta \hat T^{\nu}\rangle.$  By virtue of Eqs.~\eqref{+,-} and
\eqref{npm} this pairing is proportional to the density-density correlation function $\langle
(n^{\mu}(\mathbf r_1, \mathbf a_1) -n)(n^{\nu}(\mathbf r_2, \mathbf a_2) -n)\rangle $ for $\mu=\pm,
\nu=\pm.$
 Assuming that disks are randomly distributed over the sample we get
for these functions
\begin{align}\label{pair}
&\langle (n^{\mu}(\mathbf r_1, \mathbf a_1) -n)(n^{\nu}(\mathbf
r_2, \mathbf a_2) -n)\rangle =
\\
\nonumber &n\int d\mathbf R \delta(\mathbf r_2-\mathbf R + \mu~
\mathbf a_1) \delta(\mathbf r_1-\mathbf R + \nu~ \mathbf a_2)
\end{align}
It is natural to interpret the vector $\mathbf R$ as the position
of the center of the disk on which a double scattering occurs.
However, this interpretation is valid  for  diagram $(+,+) $ only. As shown in Fig.~\ref{fig6},
this diagram corresponds to the situation where an electron experiences two real scattering
processes on the disk placed at point $ \mathbf R .$ The physical interpretation of other diagrams
is more subtle. The diagram $(-,-)$ (see Fig.~\ref{fig9}) does not correspond to any real
scattering at point $\mathbf R.$ It just allows one to calculate correctly the probability for an
electron to pass twice the region of the size $a$ around point $\mathbf R$ without scattering.  To
interpret the diagrams $(+,-)$ and $(-,+),$ note that in the Boltzmann picture, which neglects
correlations, the following process is allowed.  An electron scatters on a disk and later on passes
through the region occupied by this disk without a scattering (see Fig.~\ref{fig8}) (analogous
consideration is valid for diagram shown in Fig.~\ref{fig7}). The diagrams $(+,-)$ and $(-,+)$
correct the Boltzmann result by subtracting the contribution of such unphysical processes.

As follows from the qualitative discussion Section~\ref{Qualitative},  the processes of ballistic
returns after a single scattering give rise to the leading (non-analytic) correction to the
Drude-Boltzmann result in zero magnetic field. However, taking into account diagrams shown in
Fig.~\ref{fig3}a (describing processes shown in Figs.~\ref{fig6}-\ref{fig7}) is not sufficient for
calculation of the low field anomaly of the MR. Actually, diagrams in Fig.~\ref{fig3}a do not
contain the "empty corridor" effect.  We will show that the correct description of the memory
effects, specific for ballistic returns, requires the renormalization of diagrams \ref{fig3}a by
diagrams \ref{fig3}b, in which the internal dashed lines contains $(-,-)$ pairings only.
Physically, the $N$'th order diagram of the type \ref{fig3}b represents the $N-1$'th order term in
the Taylor expansion of the $\exp(-nS_0)$ in the qualitative estimate \eqref{estimate}. Such
renormalization play the key role in the quantitative description of the anomalous MR.

In the next subsection we derive analytical expression for  diagrams \ref{fig3}a. Then we
generalize the calculations to account for the corridor effect and calculation of anomalous
MR.

\subsection{Non-analytical corrections to the zero-field resistance neglecting the "empty corridor" effect}

In this subsection we use the perturbative expansion Eq.~\eqref{series} to derive analytically the
leading correction to the diffusion coefficient. This correction is due to the processes described
by diagram Fig.~\ref{fig3}a. The operator expression of the correction to the Green's function
related to this diagram is given by
\\
\begin{picture}(0,30)(50,-15)
\put(250,-7){\line(1,0){68}} \put(318,-7){\line(0,1){5}} \put(250,-7){\line(0,1){5}} \put(207,0) {$
\displaystyle{ \sum_{\mu,\nu=\pm} \hat G^B  \delta \hat T^{\mu}\hat G^-\langle\hat T^+ \rangle \hat
G^-\delta \hat T^{\nu} \hat G^B,}$} \label{pic}
\end{picture}
\\
\noindent
 where the underbracket stands for the pairings of operators $\delta \hat T^{\mu}$
and $\delta \hat T^{\nu}$.
 Substituting this correction into Eq.~\eqref{D1}, and
using Eqs.~\eqref{intg},\eqref{lm} to integrate over $\mathbf r ,~
\mathbf r_0,~ \alpha,~ \alpha_0$ we get
\begin{equation}
\delta D= \frac{v_F l_{{\rm tr}}^2}{4\pi}\int d \beta_0 d \beta
M_{\beta,\beta_0} \cos(\beta-\beta_0) \label{dD}
\end{equation}
where
 \begin{align}\label{M1}
 M_{\beta, \beta_0}& = \frac{1}{\rm S} \sum_{\mu,\nu =\pm}
 \int d \mathbf r_{*} d \mathbf r_1 d\mathbf r_2
d\alpha_1 d\alpha_2d\beta_1d\beta_2
 \\
 \nonumber
 &
 \left\langle\delta T^{\mu}_{\beta, \alpha_2}(\mathbf r_2) G^{-}_{\alpha_2, \beta_2}
 (\mathbf r_2-\mathbf r_*)n \sigma(\beta_2-\beta_1)\right.
 \left. G^{-}_{\beta_1, \alpha_1}( \mathbf
 r_*-\mathbf r_1) \delta T^{\nu}_{\alpha_1, \beta_0}(\mathbf r_1)
 \right \rangle.
\end{align}
Here we used Eq. \eqref{tt} for $\langle \hat T^+ \rangle$.
\begin{figure}
\includegraphics[width=0.35\textwidth]{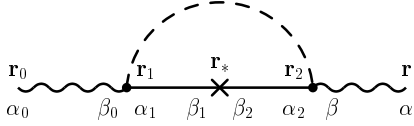}
\caption{Graphical representation of Eqs.~\eqref{dD}, \eqref{M1}} \label{fig5}
\end{figure}
Graphically, these calculations are presented in Fig.~\ref{fig5}.
 The
pairings $\langle \delta T^{\mu}_{\beta,\alpha_2}(\mathbf r_2)
\delta \hat T^{\nu}_{\alpha_1, \beta_0}(\mathbf r_1)\rangle ,$
entering Eq.~\eqref{M1}, are given in Appendix~A for four
possible combinations of $\mu, \nu$. The pairings only depend on
the difference $\mathbf r_1- \mathbf r_2$. Therefore in Eq.
\eqref{M1} one can remove the integral over $\mathbf r_*$ and the
sample area $\rm S$ in the denominator and put $\mathbf r_*=0$ in
the integrand. In other words we put the origin of the coordinate
system at the backscattering point $\mathbf r_*.$ Next we
integrate Eq.~\eqref{M1} over the angles $\alpha_1, \alpha_2, \beta_1,
\beta_2.$ This integration can be easily done taking into account
angle dependent delta functions entering in $G^{-}_{\alpha_2,
\beta_2}
 (\mathbf r_2)
 $ and $ G^{-}_{\beta_1, \alpha_1}(-\mathbf r_1) $ (see Eq.~\eqref{G-}).
 As a result we get $\alpha_1=\beta_1=\alpha[-\mathbf r_1],~
 \alpha_2=\beta_2=\alpha[\mathbf r_2]$ and
 \begin{align}\label{M3}
&M_{\beta, \beta_0} \approx n \sigma(\pi) \sum_{\mu,\nu =\pm}
 \int d \mathbf r_{1} d\mathbf r_2
\\
 &\left\langle\delta T^{\mu}_{\beta, \alpha[\mathbf r_2]}(\mathbf r_2)
 \delta T^{\nu}_{\alpha
 [ - \mathbf r_1]
 , \beta_0}(\mathbf r_1)
 \right \rangle
 \nonumber
 \frac{e^{- r_2/l}
 e^{-r_1/l}}{r_2 r_1}
\end{align}
In derivation of Eq.~\eqref{M3} we took into account that the
pairing of collision operators $\langle \delta T^{\mu}(\mathbf
r_1) \delta T^{\nu}(\mathbf r_2)\rangle$ vanishes for $\vert
\mathbf r_1 - \mathbf r_2 \vert > 2 a, $  which is clear from its
physical meaning and also can be seen from the delta functions in
the factor $J(\mathbf a, \mathbf b)$ (see Appendix  A). At the
same time both $r_1$ and $r_2$ are of the order of the mean free
path $l\gg a.$ Therefore, $\beta_2-\beta_1 =\alpha[\mathbf
r_2]-\alpha[ -\mathbf r_1] \approx \pi,$ and
$\sigma(\beta_2-\beta_1) \approx \pi.$ Physically this means that
the typical scattering angle in the processes of ballistic return are
close to $\pi.$ Next, we use Eq.~\eqref{M3} to evaluate the
contributions of different pairings $\mu, \nu.$

\subsection*{(+,+) pairing}

Consider first the process in which $\mu=+, \nu=+.$ Substitute
equations \eqref{4++} and \eqref{M3} in Eq. \eqref{dD} and
integrate over $\mathbf r_1, ~ \mathbf r_2.$ It is convenient for
our purposes to perform this integration for a fixed value of the
vector $\mathbf R$. Then the delta functions in Eq.~\eqref{J}
ensure that $\mathbf r_1$ and $\mathbf r_2$ lie on the circle of
the radius $a$ centered in the point $\mathbf R$
\begin{align}
\label{r12++} &\mathbf r_1=
 \mathbf R - \mathbf a_{\alpha[-\mathbf r_1],\beta_0 }
\\
\nonumber &\mathbf r_2= \mathbf R -\mathbf a_{ \beta,
\alpha[\mathbf r_2] }.
\end{align}
The graphical solution of these equations is shown in
Fig.~\ref{fig6} (  vectors $\mathbf n_{\beta},\mathbf n_{\beta_0}$
in this figure are the unit vectors in the directions of $\beta$
and $\beta_0$ respectively).
 Since $a$ is small ( $a
\ll l $) one has approximately
\begin{equation}\mathbf r_1 \approx \mathbf r_2
\approx \mathbf R. \label{rrR}
\end{equation}
This accuracy is sufficient for the calculation of the diagram in
Fig.~\ref{fig3}a (since we calculate it in the lowest order in
$a/l$ ). Upon integrating out the vectors $\mathbf r_1,~\mathbf r_2$
the integrand in Eq.~\eqref{dD} depends on $\mathbf R.$ Finally we
get
\begin{align}
\label{D++}
 \delta D^{ ++} =& \frac{v_F l_{\rm tr}^2 n^2
\sigma(\pi)}{4\pi } \int d \mathbf R   d\beta  d \beta_0
\cos(\beta - \beta_0)
\\
& \sigma(\beta - \alpha[\mathbf R
  ]) \sigma(\alpha[-\mathbf R ] -
\beta_0)\frac{e^{-2 R/l}}{ R^2} \nonumber
\end{align}
Introducing new variables $\varphi_0=  \beta_0 - \alpha[-\mathbf
R]$ and $\varphi_f=\alpha[\mathbf R ]- \beta$ and taking into
account that $\alpha[\mathbf R]=\alpha[-\mathbf R] +\pi$ and
$\sigma(-\varphi) =\sigma(\varphi),$ the relative correction to
the diffusion coefficient is written as
\begin{equation}
\frac {\delta D^{++}}{D_0}= - \frac{n l_{\rm tr}}{4l} \int
\frac{dR }{R} d \varphi_0 d\varphi_f ~ e^{-2R/l}
\sigma(\varphi_f)\sigma (\varphi_0)\cos(\varphi_0+ \varphi_f)
\label{dD++}
\end{equation}
>From Fig.~\ref{fig6} it is seen that $\varphi_0$, $\varphi_f$ are the scattering angles.
\subsection*{(+,-) pairing}
 Next we present calculations for $\mu=+,~\nu=-.$
   Like the calculation
 of   the diagram $(+,+),$
we substitute Eq.~\eqref{M3} and Eq.~\eqref{4+-} into
Eq.~\eqref{dD} and integrate over $\mathbf r_1,~ \mathbf r_2$ with
the use of Eq.~\eqref{J}.  The values of $\mathbf r_1$ and
$\mathbf r_2$ should be found from the following equations
\begin{align}\label{r12+-}
&\mathbf r_1= \mathbf R + \mathbf a_{\alpha[ - \mathbf r_1],
\alpha'}
\\
\nonumber &\mathbf r_2= \mathbf R - \mathbf
a_{\beta,\alpha[\mathbf r_2 ]}.
\end{align}
The graphical solution of these equation is presented in
Fig.~\ref{fig7}. After integrating out vectors $\mathbf
r_1,~\mathbf r_2$
and the angle $\beta_0,$   we get
\begin{align}
\label{D+-} &\delta D^{ +-} = - \frac{v_F l_{\rm tr}^2 n^2
\sigma(\pi)}{4\pi } \int d \mathbf R   d\beta  d \alpha'
\cos(\beta - \alpha[-\mathbf R])
\\
\nonumber
 & \sigma(\beta - \alpha[\mathbf R
  ]) \sigma(\alpha[-\mathbf R ] -
\alpha')\frac{e^{-2 R/l}}{ R^2}
\end{align}
In derivation of this equation we have used Eq.~\eqref{rrR}. It is
convenient to rewrite this equation using  as integration
variables $\varphi_0,~\varphi_f,$ which are  the scattering angles
for the process $(+,+)$. They are expressed in terms of $\alpha'$
and $\beta $ as $\varphi_0=\alpha[-\mathbf R] - \alpha',
~\varphi_f= \alpha[\mathbf R]- \beta.$
  In these variables
 Eq.~\eqref{D+-} reads
\begin{equation}
\frac {\delta D^{+-}}{D_0}= \frac{n l_{\rm tr}}{4l} \int \frac{dR
}{R} d \varphi_0 d\varphi_f ~ e^{-2R/l} \sigma(\varphi_f)\sigma
(\varphi_0)\cos( \varphi_f)
 \label{D+-1}
\end{equation}
It is seen from Eq.~\eqref{D+-1} (see also Fig.~\ref{fig7}) that
the contribution of the process $(+,-)$ can be parameterized by the
angles $\varphi_0$ and $\varphi_f,$ which are the scattering
angles for the process $(+,+).$
 Analogous calculations can be easily done for other types of
correlations.

 Summing the different contributions $\delta D=
\delta D^{++}+ \delta D^{+-}+ \delta D^{-+}+\delta D^{--}$ we get
the following expression for
 the diagram Fig.~\ref{fig3}a
\begin{align} \label{fin00}
&\frac{\delta \rho}{\rho} = -\frac{\delta D} {D}=
   \frac {n
l_\text{tr}}{4l} \text{Re} \int_a^{\infty} \frac{dR}{R}e^{-2R/l}
\int_0^{2\pi} d \varphi_0 \int_0^{2\pi} d \varphi_f
\sigma(\varphi_0) \sigma(\varphi_f)
\\
\nonumber
 &(1- e^{i \varphi_0}) (1- e^{i \varphi_f})   =
\frac{2\beta_0}{3}\ln\left(\frac{1}{2\beta_0}\right).
\end{align}
This equation is the exact expression for the non-analytic correction
\cite{dorfman,wei,wei1,hauge,peierls}
to the Drude-Boltzmann resistivity
which was qualitatively given in Eq.~\eqref{Weil}. Four terms in the product $(1-e^{i
\varphi_0})(1-e^{i \varphi_f})= 1 -e^{i\varphi_0}-e^{i\varphi_f}+
e^{i(\varphi_0 +\varphi_f)}$ correspond to four types of
correlations discussed above.

\subsection{"Empty corridor" effects on the zero-field
resistance} In this subsection we use the perturbation theory for the quantitative derivation of
the effect of the "empty corridor" discussed in section \ref{Qualitative}.

Eq.\eqref{fin00} takes into account one pairing of operators $\delta T^{\mu}$ and $\delta T^{\nu}
.$ The terms containing  $N$ pairings ($N$ dashed lines) are typically small as $\beta_0^N.$
However, there is a series of diagrams, shown in Fig.~\ref{fig3}b, whose contribution is of the
order $\beta_0$ (see Ref.~\onlinecite{wei1}). The internal dashed lines in this series only contain
pairings $\langle \delta \hat T^- \delta \hat T^- \rangle.$ Below we show that the series
Fig.~\ref{fig3}b accounts for the effect of the ''empty corridor''. More precisely, we prove that
the $N-$th order term in this series corresponds to $N-1$ term in the Taylor expansion of the
$\exp(n S_0)$ in Eq.~\eqref{p}.

\begin{figure}
\includegraphics[width=0.60\textwidth]{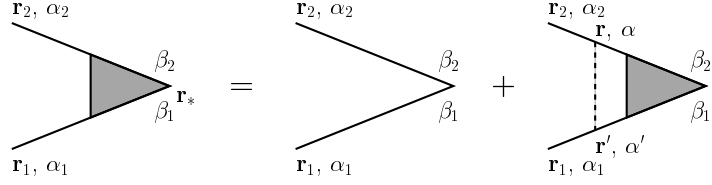}
\caption{Dyson equation, describing the  renormalization of
backscattering events by the "empty corridor" effect. $N-$th order
term in the these stair-like diagrams corresponds to the
$(N-1)-$th term in the Taylor expansion of $\exp(nS_0)$ in
Eq.~\eqref{p}. } \label{fig10}
\end{figure}

\begin{figure}
\includegraphics[width=0.35\textwidth]{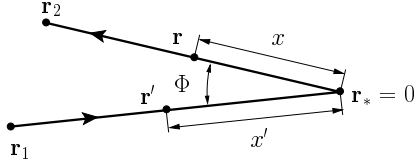}
\caption{Pairings $\langle \delta \hat T^- \delta \hat T^-
\rangle$ do not change the electron trajectory. Point $\mathbf r$
(see Fig.~\ref{fig10}) lies on the line segment connecting point
$\mathbf r_1$ and the origin and point $\mathbf r^{\prime}$ lies
on the line segment connecting point $\mathbf r_2$ and the origin}
\label{fig11}.
\end{figure}

\begin{figure}
\includegraphics[width=0.45\textwidth]{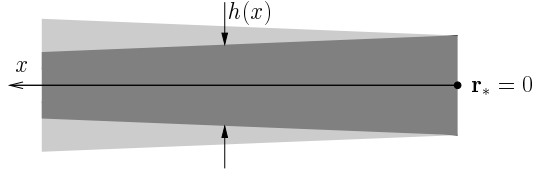}
\caption{Width of the overlap region as a function of a distance
from the origin} \label{fig12}
\end{figure}

The addition  of the diagram  Fig.~\ref{fig3}b to the diagram
Fig.~\ref{fig3}a leads to the following replacement in
Eq.~\eqref{M1}
\begin{equation}
    G^-_{\alpha_2,\beta_2}(\mathbf r_2 )
    G^-_{\beta_1,\alpha_1}(- \mathbf r_1) \to
    K^{\alpha_1,\alpha_2}_{\beta_1,\beta_2}(\mathbf r_1,\mathbf r_2),
\label{line}
\end{equation}
where the function $K^{\alpha_1,\alpha_2}_{\beta_1,\beta_2}(\mathbf
r_1,\mathbf r_2)$
 can be found from
    the Dyson equation (See Fig.~\ref{fig10}).
     Since the operator $\delta
    T^-$ does not change the velocity angle, one can search the
    solution of the Dyson equation in the form
    \begin{align}\label{K}
    &K^{\alpha_1,\alpha_2}_{\beta_1,\beta_2}(\mathbf r_1,\mathbf r_2)=
\\
\nonumber
    &G^-_{\alpha_2,\beta_2}(\mathbf r_2 )
    G^-_{\beta_1,\alpha_1}(- \mathbf r_1) \Gamma(r_1,r_2,\Phi),
\end{align}
where  $\Phi$ is the difference between the backscattering angle
and $\pi$
(See Fig.~(\ref{fig11})). While substituting Eq.~\eqref{K} in the
Dyson equation (which is shown graphically in Fig.(\ref{fig10}),
integrating over angles $\alpha,\alpha'$, and using  the identity

\begin{equation}
\delta(\alpha[\mathbf r_2 -\mathbf r] -\alpha[\mathbf
r])=\frac{\vert\mathbf r_2 -\mathbf r
\vert}{r_2}\delta(\alpha[\mathbf r] -\alpha[\mathbf
r_2])\theta(r_2-r) \label{identity}
\end{equation}
(and analogous identity for vectors $\mathbf r', \mathbf r_1$ )we
get
\begin{align}
&\Gamma(r_1,r_2,\Phi)=1 +
\\
\nonumber &\int_{r<r_2} \int_{r'<r_1} d\mathbf r   d\mathbf
r^{\prime} \frac{\delta(\alpha[\mathbf r]- \alpha[\mathbf r_2]
)\delta(\alpha[- \mathbf r']-\alpha[-\mathbf r_1])}{r r'}\left
\langle \delta T^-_{\alpha[\mathbf r_2 ]}(\mathbf r) \delta T^-_{
\alpha[-\mathbf r_1]} (\mathbf r^{\prime})\right \rangle
\Gamma(r,r^{\prime},\Phi), \label{Psi0}
\end{align}
where $\delta T^-_{\alpha}(\mathbf r)$ are defined in the Appendix~ \ref{ApA}. Further calculations
can be done in two different ways. It is useful from methodological point of view to discuss both
of them. One way is to first integrate Eq. \eqref{Psi0} over the angles of the vectors $\mathbf r$
and $\mathbf r'$ (see Fig.~\ref{fig10}). The delta-functions entering in Eq.~\eqref{Psi0} ensure
that point $\mathbf r$ (see Figs.~\ref{fig10},\ref{fig11}) lies on the line segment connecting
point $\mathbf r_1$ and the origin and point $\mathbf r^{\prime}$ lies  on the line segment
connecting point $\mathbf r_2$ and the origin.
This allows to reduce the Dyson equation to the closed relation
for function  $\Gamma(r_1,r_2,\Phi)$
\begin{align}
&\Gamma(r_1,r_2,\Phi)=1 +
\\
\nonumber &\int_0^{r_1}dr' \int_0^{r_2} dr \left \langle \delta
T^-_{\alpha[\mathbf r_2 ]}(\mathbf r) \delta T^-_{ \alpha[-\mathbf
r_1]} (\mathbf r^{\prime})\right \rangle \Gamma(r,r^{\prime},\Phi)
\label{Psi}
\end{align}
where the pairing
\begin{equation}
\left \langle \delta T^-_{\alpha[\mathbf r_2 ]}(\mathbf r) \delta
T^-_{ \alpha[-\mathbf r_1]} (\mathbf r^{\prime})\right \rangle
\approx
h(r,\Phi)\delta(r-r^{\prime}), \label{dTmap}
\end{equation}
is calculated in Appendix \ref{ApB}. Here
\begin{equation}
h(r, \Phi) = \left(2a - \left\vert \Phi   \right\vert
r\right) \theta \left(2a - \left\vert \Phi  \right\vert r\right)
\label{hdef}
\end{equation}
 is shown in Fig.\ref{fig12}.
  As a result we
have
\begin{align}
&\Gamma(r_1,r_2,\Phi)=1 +
\\
\nonumber &n\int_0^{\min[r_1,r_2]} dr h(r,\Phi)\Gamma(r,r,\Phi)
\label{Psi1}
\end{align}
This equation has an evident solution
\begin{equation}
\Gamma(r_1,r_2,\Phi)=\exp \left(n\int_0^{\min[r_1,r_2]} dr
h(r,\Phi)\right).
\label{Psi1}
\end{equation}

Eq.~\eqref{Psi1} may be derived in an alternative way. This way
allows to understand on the formal basis why one should
renormalize the diagrams in Fig.~\ref{fig3}a  by  the $\langle
\delta T^- \delta T^-\rangle$ pairings only. Let us integrate in
Eq.~\eqref{Psi0} over $\mathbf r, \mathbf r',$ using Eq.~
\eqref{4--}. Doing this, we get
\begin{align}
&\Gamma(r_1,r_2,\Phi)=1 +
\nonumber \\
&n \int  d\alpha' d\alpha'' \int_{R<\min[r_1,r_2]} d\mathbf R \frac{
\delta(\alpha[\mathbf R +\mathbf a_{\alpha[\mathbf r_2],\alpha'}]-
\alpha[\mathbf r_2] ) \delta(\alpha[-\mathbf R -\mathbf
a_{\alpha[- \mathbf r_1],\alpha''}]- \alpha[-\mathbf r_1] )}{R^2}
\nonumber \\
& \sigma(\alpha[\mathbf r_2] -\alpha') \sigma(\alpha[\mathbf r_1]
-\alpha'')\Gamma(R,R,\Phi) .
\label{PsiPsi}
\end{align}
Here we took into account that $\Gamma(r,r',\Phi)$ is a slowly changing function  of $r,r'$ (it
changes on a scale of the order of $l$). Since Eq.~\eqref{4--} provide $|\mathbf r -\mathbf R| \sim
|\mathbf r' -\mathbf R| \sim a$, we put $\Gamma(r,r',\Phi) \approx \Gamma(R,R,\Phi).$ Next we make
use of the identity
\begin{equation}
\int d \alpha' \sigma(\alpha -\alpha') \delta(\alpha[\mathbf R +\mathbf
a_{\alpha,\alpha'} ]-\alpha ) \approx R~\theta(a/R -\vert \alpha
-\alpha[\mathbf R]\vert)
 \label{--cross}
\end{equation}
to integrate over $\alpha', \alpha''.$ After this integration we get
\begin{align}
&\Gamma(r_1,r_2,\Phi)=1 +
\nonumber \\
&n  \int^{\min[r_1,r_2]}_0 R d  R d\alpha[\mathbf R] \theta(a/R
-\vert \alpha[\mathbf r_2] -\alpha[\mathbf R]\vert)\theta(a/R
-\vert \alpha[\mathbf r_1] -\alpha[\mathbf R]\vert)
\Gamma(R,R^{\prime},\Phi) \label{PsiPsiPsi}
\end{align}
Integration over $\alpha[\mathbf R]$ leads again to
Eq.~\eqref{Psi1}.

>From Eq.~\eqref{rrR} it follows that we need to know $\Gamma$ for $r_1 \approx r_2 \approx R.$
Using Eq.~\eqref{Psi1} we find
\begin{equation}
\Gamma(R,R, \Phi)=\exp \left(n\int_0^{R} dr h(r,\Phi)\right) = e^{n
S_0(R,\Phi)} \label{psifin}
\end{equation}
where $ S_0(R,\Phi)$ is the overlap between two corridors, given by Eq.~\eqref{s00}. The $n$-th
order term  in the Taylor expansion of the exponential \eqref{psifin} corresponds to a diagram in
Fig.~\ref{fig3}b with $n+1$ dashed lines. Indeed, the $n$-th order term in the Taylor expansion
contains $n$ integration over coordinates of $n$ scatterings of the type $-,-$. What remains to do
to get the resistivity correction is to express $\Phi$ in Eq.~\eqref{psifin} via angles $\varphi_0$
,$\varphi_f.$ To do this, the precision Eq.~\eqref{rrR} is not sufficient. More specifically, we
will need to know the angle between vectors $\mathbf r_1$ and $\mathbf r_2$ to the order $a/l.$ For
this purpose it will be sufficient to make replacements $\alpha[ - \mathbf r_1]\approx \alpha[ -
\mathbf R]$  and $\alpha[\mathbf r_2] \approx \alpha[\mathbf R ]$ in the arguments of vectors
$\mathbf a$ in Eq.~\eqref{r12++}.
Then the angle $\Phi$ is calculated as (see Figs.~\ref{fig6},\ref{fig9},\ref{fig8},\ref{fig7})
\begin{equation}
\Phi \approx \frac{a}{R}~[\cos(\varphi_0/2) + \cos(\varphi_f/2)]
 \label{Phi}
\end{equation}
This equation is valid for $0 < \varphi_0 <2 \pi $, $0<\varphi_f <
2\pi$ (since $\cos(\varphi/2)$ is not periodic with $2\pi$ one
should specify the integration limits). It worth noting that
Eq.~\eqref{Phi} ensures that the argument of the $\theta$ function
in Eq.~\eqref{hdef} is positive and the overlap width and overlap
area are given  by
\begin{align}
&h(r, \Phi) = 2a - \left\vert \Phi   \right\vert r \nonumber \\
&S_0(R,\Phi)=\int_0^R h(r, \Phi) dr =2aR-\left\vert \Phi
\right\vert R^2/2\label{h1}
\end{align}

Summing the diagrams Fig.~\ref{fig3}b together with
Fig.~\ref{fig3}a, one gets an exact equation
\begin{align} &\frac{ \delta \rho}{\rho}= \frac {n
l_\text{tr}}{4l} \text{Re} \int_a^{\infty} \frac{dR}{R}e^{-2R/l}
\int_0^{2\pi} d \varphi_0 \int_0^{2\pi} d \varphi_f
\sigma(\varphi_0) \sigma(\varphi_f) \nonumber \\ &(1- e^{i
\varphi_0}) (1- e^{i \varphi_f}) e^{ nS_0 (R, \Phi)} =
\frac{2\beta_0}{3}\ln \left (\frac{C}{2\beta_0 }\right),
\label{fin1} \end{align}
instead of qualitative estimate Eq.~\eqref{estimate}.

Here  $C \approx 1.8$ is  a numerical coefficient. Thus, addition of the series Fig.~\ref{fig3}b to
Fig.~\ref{fig3}a leads to the following renormalization: $ \ln(1/2\beta_0) \to \ln(C/2\beta_0).$
%

\subsection{Estimate for neglected diagrams  }
In this subsection we use the derivation presented in the previous subsection to show how to select
relevant diagrams. The derivation was based on identity \eqref{--cross}. The left hand side of this
equation is proportional to the impurity cross section $\sigma(\alpha -\alpha')\sim a.$ However,
the right hand side is parameterically larger $\sim R,$ provided that $|\alpha -\alpha[\mathbf
R]|<a/R .$ One can show that the function similar to the one  on the right hand side
\eqref{--cross} arises  each time when one of the index $\mu$ or $\nu$ in the pairing $\langle
\delta T^{\mu} \delta T^{\nu}\rangle$ is equal to $(-).$ (for a pairing $\langle \delta T^{-}
\delta T^{-}\rangle$ we have two functions of the type \eqref{--cross}, see Eq.~\eqref{PsiPsiPsi}).
For the case when $\mu$ or $\nu$ equals to $(+)$ we have a differential cross section
$\sigma(\alpha -\alpha[\mathbf R]) \sim a $ instead of function \eqref{--cross}. Now we are ready
to estimate different types of diagrams. Insertion of one additional $(-,-)$ pairing into a diagram
in Figs.~\ref{fig6}-\ref{fig7} gives two additional Green's functions and, consequently, multiplier
$\sim 1/l^2,$ and two functions of the type \eqref{--cross} giving a multiplier $\sim l^2.$ One
should also multiply on disk concentration $n$ and integrate over $d \mathbf R.$ Due to $\theta $
functions in Eq.~\eqref{--cross} the integration area $d\mathbf R$ is of the order $l a.$ Combining
all the multipliers together we have $nla \sim 1. $ Therefore, addition of $(-,-)$ pairing does not
lead to any smallness. In contrast to this, insertion of $(-,+)$ pairing leads to smallness $\sim
a/l.$ Indeed, the only difference from the case of $(-,-)$ pairing is that one should replace one
of the function \eqref{--cross} by the corresponding cross section. This leads to the change of one
of the multiplier of the order $l$ by multiplier of the order $a$. Following this line of reasoning
one could conclude that insertion of $(+,+)$ pairing leads to the smallness of the order of
$(a/l)^2.$ Actually, the smallness arising from insertion of $(+,+)$ pairing is of the order of
$a/l.$ Indeed, while the replacement of each of two function of the type $\eqref{--cross}$ by
corresponding cross section leads to the relative smallness of the order $(a/l)^2,$ the integration
area in this case is not restricted by angle dependent $\theta$ functions. As a result, the
integration area,
 $d\mathbf R \sim l^2,$ is larger  by a parameter $l/a$ compared to
the case of $(-,-)$ pairing. Note that estimates presented above do not work for the diagrams with
one dashed line Figs.~\ref{fig6}-\ref{fig7}. In this case one should integrate over initial and
final scattering angles. One can easily see that integration of Eq.~$\eqref{--cross}$ over angle
$\alpha$ leads to the multiplier of the order of $a$. This implies that in this case all four
pairings, $(++)$,$(+-)$,$(-+)$, and$ (--)$ are of the same order, of $a/l$ (see Eq.~\eqref{fin00}).
Note that the higher order diagrams which are small in the parameter $\beta_0$ may turn out to be
relevant for the MR at very low magnetic field (see Section \ref{lyap}).

\subsection{Anomalous magnetoresistance \label{MR}}
\begin{figure}
\includegraphics[width=0.40\textwidth]{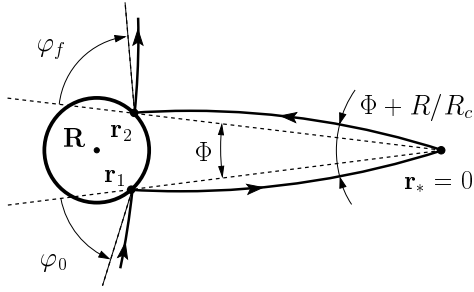}
\caption{The magnetic field changes the backscattering angle
$\phi=\Phi +R/R_c.$ The dashed (solid) line in  represents
electron trajectory for $B=0$ ($B\neq 0$).  The electron
trajectory for $B\neq 0$ is parameterized by the scattering angles
$B= 0$} \label{fig13}
\end{figure}

In this subsection we generalize our calculations for the $B \neq 0$  case assuming that $B$ is
small ($\beta \ll 1).$ The main contribution in this case still comes from diagrams in
Figs.~\ref{fig3}a,b. Consider for example diagram $(++)$. Let us compare the process of double
scattering described by this diagram for $B \neq 0$ (see Fig.~\ref{fig13}) with the same process
for $B=0$ (see Fig.~\ref{fig6}). For fixed points $\mathbf r_1,$ $\mathbf r_2$ and $\mathbf r_*=0$,
one can see the following differences. First, the scattering angles $\varphi_0$ and $\varphi_f$
acquire small corrections of the order of  $R/R_c
 \sim l/R_c = \beta \ll 1.$ Second, the parts of the electron
trajectory corresponding to free ends of the picture become
curved. The backscattering angle $\phi$  increases by the value
$R/R_c$
\begin{equation}
\phi =\Phi +R/R_c,
 \label{phi}
\end{equation}
where $\Phi$ is the value of backscattering angle for $B=0$ given
by Eq.~\eqref{Phi}. Finally, the overlap area of the corridors
changes because the trajectories  become curved (see
Fig.~\ref{fig13}).
 The corrections to $\varphi_0$ and $\varphi_f$
lead to small relative corrections to the resistivity of the order
of $\beta_0 \beta^2$ and can be neglected. The same reason allows one
to neglect the curvature of the incoming and outgoing parts of the
trajectory. Therefore,the only relevant difference is the change
of the overlap area of the corridors. The solution of the Dyson
equation is analogous to the  $B=0$ case. The points of
intermediate integration $\mathbf r$ and $\mathbf r'$ lie now at
the segments of cyclotron circles (from $\mathbf r_1 $ to $\mathbf
r_*$ and from $\mathbf r_* $ to $\mathbf r_2$). The pairing of two
operators $\delta \hat T^-$ is still given by Eq.~\eqref{dTmap}
with the replacement $h(r) \to h_B(r).$  For $\beta,\beta_0 \ll 1$
the overlap width is calculated as
\begin{equation}
h_B(x) \approx \left(2a - \left\vert \phi r - r^2/R_c
\right\vert\right) \theta \left(2a - \left\vert \phi r - r^2/R_c
\right\vert \right) , \label{hb}
\end{equation}
where $\theta$ is the Heaviside step function.
 Therefore,  the only difference from
Eq.~\eqref{fin1} is that one should replace $S_0 \to S_B,$ where
 the overlap area is given by
\begin{equation}
 S_B (R,
\phi) = \int_0^R d r h_B(r) , \label{sb}
\end{equation}
The value of $\delta \rho_{xx}/\rho$ is obtained from
Eq.~\eqref{fin1} by replacing $e^{n S_0}$ to $e^ {n S_B} - e^{n
S_0}$
\begin{align} &\frac{ \delta \rho_{\rm xx}}{\rho}= \frac {n
l_\text{tr}}{4l} \text{Re} \int_a^{\infty} \frac{dR}{R}e^{-2R/l}
\int_0^{2\pi} d \varphi_0 \int_0^{2\pi} d \varphi_f
\sigma(\varphi_0) \sigma(\varphi_f) \nonumber \\ &(1- e^{i
\varphi_0}) (1- e^{i \varphi_f})\left(e^{ nS_B (R, \Phi)}- e^{
nS_0 (R, \Phi)}\right), \label{fin2} \end{align}
Introducing dimensionless variables $T=R/l, \ z= \beta/\beta_0$ we
get
 Eq.~\eqref{1}, where
function $f(z)$ is given by
\begin{align}
&f(z)= \frac{3}{32} \int_0^{\infty}\frac{dT}{T}
e^{-2T} \int_0^{2\pi} d \varphi_0 \int_0^{2\pi} d
\varphi_f
\nonumber \\ & \cos\left(\frac{\varphi_0
+\varphi_f}{2}\right)
\sin^2\left(\frac{\varphi_0}{2}\right) \sin^2
\left(\frac{\varphi_f}{2}\right) \left( e^{ s_z} -
e^{ s_0}\right).  \label{f(z)} \end{align}
Here
\begin{align} &s_z= \int_0^{T} dt \left(1-\left\vert
\zeta t -\frac{z t^2}{2} \right\vert \right) \theta
\left(1- \left\vert \zeta t -\frac{z t^2}{2}
\right\vert \right),
\nonumber \\ &\zeta=\frac{\cos(\varphi_0/2)
+\cos(\varphi_f/2)}{2T} +\frac{z T}{2}, \quad s_0=
s_{z \to 0}.  \label{s_z} \end{align}
Function $f(z)$ has the following asymptotics
\begin{equation} f(z)= \left \{ \begin{array}{ll}
0.32 z^2 & \text{for $z \to 0$}
\\
0.39 - 1.3 /\sqrt{z} & \text{for $ z \to \infty$}.
\end{array} \right.
\label{asimpt} \end{equation}
In the interval $0.05 \lesssim z
\lesssim 2,$  $f(z)$ can be well approximated by linear function
\begin{equation}
f(z)\approx 0.032\ (z -0.04) ~~~\text{for $ 0.05 \lesssim z
\lesssim 2$}. \label{linear}
\end{equation}

Next we discuss the parabolic asymptotics more carefully. We will
show that in this asymptotic region there are two contributions of different signs
to the magnetoresistance: a negative contribution coming from the trajectories
with very small $\Phi$ such that $|\Phi| \lesssim \beta,$ and
a positive contribution coming from larger angles. This
considerations will be used in the next section, discussing
trajectories with the long Lyapunov region.

First, we write the difference of two exponents $e^
{n S_B} - e^{n S_0}$ as follows $(e^ {n (S_B-S_0)} - 1)e^{n S_0}$
and expand the equation in the bracket in the Taylor expansion up
to the second order
\begin{align}
&e^ {n S_B} - e^{n S_0} \approx e^{n S_0}\left( n \delta S +
\frac{n^2 \delta S^2}{2}\right),\label{exp} \\
&\delta S = S_B-S_0. \nonumber
\end{align}
  We
consider the case $z \ll 1$ ($\beta \ll \beta_0$). One can easily see that in this case the
expression $2a - \left\vert \phi r - r^2/R_c \right\vert,$ entering in the argument of $\theta$
function in the Eq.~\eqref{hb} is positive at all values of $\Phi$ (we take into account
Eq.~\eqref{Phi} and have in mind that $r\lesssim l$). Therefore, the difference $\delta
S(\Phi)=S_B(\Phi)-S_0(\Phi)$ is expressed as
\begin{align}
\delta S(\Phi)&=\int_0^R dr ~r(|\Phi|-|\Phi+(R-r)/R_c|) \nonumber \\
&= \left \{ \begin{array}{ll} -\frac{R^3}{6R_c} & \text{for $\Phi>
0$}
\\
+\frac{R^3}{6R_c}-\frac{(R+R_c\Phi)^3}{3R_c}~\theta(R+R_c\Phi) &
\text{for $ \Phi < 0$}.
\end{array} \right. \label{dS}
\end{align}
To sum  the contributions of different  electron trajectories, we take into account that the time
reversed trajectories  have the same statistical weight. Indeed, as seen from
Figs.~(\ref{fig6}-\ref{fig7}),(\ref{fig13}), time reversion correspond to the change $\varphi_0 \to
2\pi-\varphi_f,~~\varphi_f \to 2\pi-\varphi_0.$ This transformation does not affect the factor
 $\sigma(\varphi_0) \sigma(\varphi_f) (1- e^{i \varphi_0}) (1- e^{i \varphi_f}),$ entering
Eq.~\eqref{fin2}. At the same time, the angle $\Phi$ changes to $-\Phi$ under this transformation
(see Eq.~\eqref{Phi}). From Eq.~\eqref{dS} we find the variation of the overlap area averaged over
two time reversed trajectories
\begin{equation}
n\frac{\delta S(\Phi) +\delta S(-\Phi)}{2}=
-\frac{n(R-R_c|\Phi|)^3}{6R_c}~\theta(R-R_c|\Phi|).
\label{smallPhi}
\end{equation}
This expression is of the order of $\beta/\beta_0$ for $|\Phi|<R/R_c
\sim \beta $ and is equal to zero for larger angles. We conclude
therefore, that time reversed contributions do not cancel  only in
the region of small angles $\Phi \lesssim \beta$. Since the total
variation of $\Phi$ is much larger, of the order of $\beta_0$ (see
Eq.~\eqref{Phi}), we can replace the expression in the right hand
side of Eq.~\eqref{smallPhi} by $\delta$-function, writing $n(\delta
S(\Phi) +\delta S(-\Phi))/2 \approx
-(\beta^2/24\beta_0)(R^4/l^4)\delta(\Phi)$. Keeping the leading
terms of the order of $\beta^2$ only, we obtain
\begin{equation}
\frac{e^{n\delta S(\Phi)} +e^{n\delta S(-\Phi)}}{2} -1 \approx
-\frac{\beta^2}{24 \beta_0}\frac{R^4}{l^4}~ \delta(\Phi) +
\frac{\beta^2}{(24 \beta_0)^2} \frac{R^6}{l^6}
\label{deltafunction}
\end{equation}
The second term in this equation comes from the averaging of the
quadratic term in the exponent expansion $n^2\delta S^2/2$ over
two time reversed trajectories. This term is positive and partly
compensates the negative contribution of small angles.
 Using Eqs.~\eqref{Phi},\eqref{h1},\eqref{fin2},  after
some algebra we get the low-field asymptotic of MR as the sum of the negative and positive
contributions discussed above
\begin{equation}
\frac{\delta\rho_{xx}}{\rho} = -\frac{\beta^2}{2\beta_0} + A~
\frac{\beta^2}{\beta_0} \approx -0.32
\frac{\beta^2}{\beta_0},\label{2contr}
\end{equation}
where numerical coefficient  $A$ is  given by
\begin{equation}
A=320\int_0^{\pi} \int_0^{\pi}d\alpha d\beta \frac{\sin^2 \alpha
\sin^2\beta ~\cos(\alpha +\beta)}{(4+ \cos \alpha -\cos \beta)^6}
\approx 0.18\label{A}
\end{equation}
\section{Contribution of  trajectories
with long Lyapunov region to magnetoresistance  \label{lyap}} 
\begin{figure}
\includegraphics[width=0.45\textwidth]{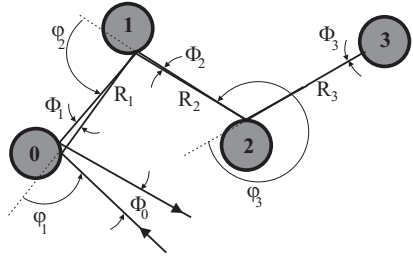}
\caption{Electron trajectory with a  Lyapunov region, containing  $N=3$ correlated links.  }
\label{fig14}
\end{figure}

\begin{figure}
\includegraphics[width=0.35\textwidth]{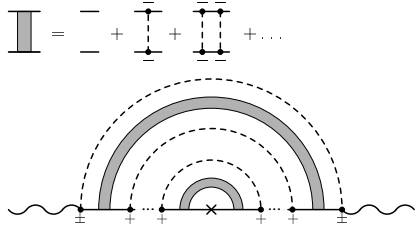}
\caption{The diagrams corresponding to electron trajectories
 with long Lyapunov regions  }
\label{fig15}
\end{figure}

The equations derived in the previous
section give the contribution to the MR related to the processes shown in
Figs.~(\ref{fig6}-\ref{fig7}), the parabolic asymptotics \eqref{2contr} starting to work when
$\beta$ becomes smaller than $\beta_0$. Such processes are related to the correlations specific for
returns to the initial point after one scattering. As we show at this section for very low magnetic
fields, $\beta \ll \beta_0^2,$ other correlations come into play. Specifically, we consider the
contribution to the MR of the trajectories containing  long Lyapunov regions. Such trajectories
consist of the direct and the return paths and involve real double scatterings on some number of
disks as shown in Fig.~\ref{fig14}. The divergency between the direct and the return paths is
characterized by the Lyapunov length. We will call such trajectories "Lyapunov trajectories". In
the diagrammatic series they are presented by the sum of the diagrams shown in Fig.~\ref{fig15}.
Just as in the case discussed above, four different pairings are allowed at the ends of external
dashed lines: $(-,-)$, $(+,-)$, $(-,+)$ and $(+,+).$ However, in contrast to the diagrams shown in
Fig.~\ref{fig3}, internal lines of "Lyapunov diagrams" contain pairings of $(+,+)$ type as well as
of $(-,-)$ type. Physically, $(+,+)$ pairings corresponds to real double scatterings in the
Lyapunov region. We will count such diagrams by the number $N$ of correlated links in the Lyapunov
region. The Lyapunov trajectory shown in Fig.~\ref{fig14} corresponds to $N=3.$ The diagrams in
Figs.~\ref{fig3} discussed in Section~\ref{calc}, present a particular case of diagrams with
Lyapunov region, corresponding to $N=1.$ The correlated links of any Lyapunov trajectory are
renormalized by the pairings $(-,-)$  as was discussed already for $N=1.$   The contribution of any
diagram of such type to the resistivity is small as $\beta_0^N$. However, as will be shown below,
these diagrams have a sharp dependence on the magnetic field at very small  fields. We will show
that for $\beta<\beta_0^N$ the diagram of the N-th order gives a contribution  to the parabolic MR
of the same order as the contribution of the diagrams with $N=1$ already calculated above.

Consider the  Lyapunov trajectory with $N$ correlated links.
Denote by $\Phi_0, \Phi_1, \dots, \Phi_N$ the angles between the
segments of the direct and the return paths and by $\varphi_0,
\dots, \varphi_{N-1}$ the scattering angles between successive
correlated links as shown in Fig.~\ref{fig14}.
  The contribution of
such a process has a sharp dependence on the magnetic field due
to the magnetic field dependence of the overlap of the corridors
surrounding the direct and the return path. This dependence is
different for different segments of the trajectory due to the
difference of the angles $\Phi_n.$ Indeed as we have seen in the
previous section if $\Phi$ is the typical angle between the direct
and the return paths then the characteristic scale for the
magnetic field dependence of the overlap of the corridors is
$\beta\sim\Phi.$ As one can see from Eq.~\eqref{Phi}, the typical
value of the angle $\Phi_n$ is of the order $\beta_0^n.$ The
smallest angle $\Phi_N\sim \beta_0^N$ corresponds to the last
segment of the trajectory. Therefore it is this segment that
should lead to the sharpest dependence of the resistivity on the
magnetic field. Consider the contribution of the last segment at
small magnetic field $\beta\ll\beta_0^N.$ For typical trajectories
with $\Phi_N\sim\beta_0^N$ there is a cancellation of the
contributions of the time reversed paths to the MR (see discussion
in the previous section). However, for a small fraction of the
trajectories with $\Phi_N \lesssim\beta$ the contribution of the
time reversed paths does not cancel and is proportional to the
change of the overlap area $ n (\delta S(\Phi_N) + \delta
S(-\Phi_N))/2 \sim \beta/\beta_0.$ The phase space of such
trajectories is proportional to $\beta/\beta_0^N.$ Thus, the
contribution of the Lyapunov trajectory with $N$ links to the
resistivity in the region of small magnetic field is given by
\begin{equation}
\frac{\delta{\rho^N_{xx}}}{\rho}\sim -\beta_0^N
\frac{\beta}{\beta_0} \frac{\beta}{\beta_0^N}\sim
-\frac{\beta^2}{\beta_0}, \quad {\rm for} \quad \beta<\beta_0^N
\label{Lyapparab}
\end{equation}
where the factor $\beta_0^N$ is due to the $N$-th order of the
corresponding diagram. As we show below, the coefficient in this
equation does not depend on the diagram order. It equals $1/2,$
thus coinciding with the coefficient in the small angles
contribution to parabolic asymptotic for $N=1$ (see
Eq.~\eqref{2contr}). It worth noting  that  only in the case
$N=1,$ one should take into account the contribution of large
angles, presented by the second term in Eq.~\eqref{2contr}. For $N
>1$ such contribution is parametrically small.

Before presenting rigorous derivation of above statements, we
consider qualitatively the contribution of $N$-th order diagram
for larger magnetic fields $\beta>\beta_0^N.$ At such fields the
first order contribution of the time reversed paths to the
corridor effect does not cancel for
 trajectories with any values of $\Phi$ up to the maximal value $\Phi \sim \beta_0^N$.
 Thus, the factor $\beta/\beta_0^N,$ which counts
the phase space of relevant trajectories saturates at some
constant value of the order of unity and one has
\begin{equation}
\frac{\delta \rho^N_{xx}}{\rho}\sim -\beta \beta_0^{N-1} \quad
\mathrm{for} \quad \beta>\beta_0^N \label{Lyaplinear}
\end{equation}

Consider now the MR in the interval
$\beta_0^{N+1}<\beta<\beta_0^{N}.$ In this interval the Lyapunov
trajectories with the number of double scatterings smaller than
$N$ give a parabolic contribution Eq.~\eqref{Lyapparab} to the MR.
The trajectories with the number of double scatterings bigger than
$n$ give a linear contribution Eq.~\eqref{Lyaplinear}, the main
contribution $\beta_0^{N}\beta$ coming from the trajectories with
$N+1$ double scattering. As a result one has
\begin{equation}
\frac{\delta \rho_{xx}}{\rho}\approx - \left(\frac {N}{2} -A
\right)\frac{\beta^2}{\beta_0}- C_N\beta_0^N \beta, \quad
\mathrm{for} \quad \beta_0^{N+1}<\beta<\beta_0^{N}.
\label{LyapQual}
\end{equation}
Here $C_N \sim 1$ and $A$ is given by Eq.~\eqref{A}.  For $N\gg 1$ the
second term in the right hand side of Eq.~\eqref{LyapQual} can be
neglected and we get finally
\begin{equation}
\frac{\delta \rho_{xx}}{\rho}\approx - \frac {N}{2}
\frac{\beta^2}{\beta_0}, \quad \mathrm{for} \quad
\beta_0^{N+1}<\beta<\beta_0^{N},\qquad N \gg 1. \label{Lyapfin}
\end{equation}
Equation Eq.~\eqref{LyapQual} indicates that at very low magnetic
fields, which correspond to large $N\sim \ln\beta/\ln\beta_0$ the
results of section \ref{MR} become incorrect and logarithmic
renormalization of the parabolic MR occurs
\begin{equation}
\frac{\delta \rho_{xx}}{\rho}\approx -\frac{\ln \beta}{2\ln
\beta_0} \frac{\beta^2}{\beta_0} \label{ln}
\end{equation}


Next we present a rigorous derivation of the contribution of the
 the $N$-th order  Lyapunov trajectory for $\beta \ll \beta_0^N$. As follows
 from qualitative considerations presented above,  $\Phi_n \ll 1$ for any $0 \leq n \leq N.$
Indeed,  $\Phi_n \sim \beta_0^n \Phi_0.$  Since the minimal angle,
$\Phi_N$ is on the order of $\beta,$ the maximal angle is small
compared to unity, $\Phi_0 \sim \beta/\beta_0^N \ll 1.$ Therefore,
 the direct and return paths of the relevant  Lyapunov
trajectories are very close to each other. This allows us to
characterize a scattering between $n-1$ and $n$ links of the
trajectory by one scattering angle $\varphi_n$ instead of two
different scattering angles for direct and return paths. The
dependence on the magnetic field comes from the last link of the
trajectory. Expanding $e^{n (S_B^N-S_0^N)}-1,$ like it was done in
the previous section, we only keep  the first term
\begin{equation}
n\frac{\delta S(\Phi_N) +\delta S(-\Phi_N)}{2}\approx
-\frac{\beta^2}{24 \beta_0}\frac{R_N^4}{l^4}~ \delta(\Phi_N).
\label{smallPhiN}
\end{equation}
To write down the analytical expression for $N$-th order
contribution we should take into account for combinations of
$(\pm,\pm)$ at the ends of diagram in Fig.~\eqref{fig15}. This
correspond to four types of scattering on the impurity $0$ in the
Fig.~\eqref{fig14} and leads to appearing of the factor $(1-e^{i
\varphi_1})(1-e^{i \varphi_1'})=|1-e^{i \varphi_1}|^2.$ Here we
took into account that direct and return paths coincides in the
first approximation and, as a consequence,  scattering angles
$\varphi_1,~\varphi_1'$ for direct and return paths are related to
each other as follows $\varphi_1=2\pi-\varphi_1'.$ The analytical
expression for MR is given by
\begin{equation}
\frac{\delta\rho_{xx}^N}{\rho}=-\frac{n^N l_{\mathrm tr}}{4 l}\int
d\varphi_{1} d\Phi_{0} d^2 \mathbf R_1\dots d^2 \mathbf R_{N}
\vert 1-e^{i\varphi_{1}}\vert^2
\frac{\sigma^2(\varphi_1)e^{-R_1/l}}{R_{1}^2}\dots
\frac{\sigma^2(\varphi_{N})e^{-R_N/l}}{R_{N}^2} \frac{\beta^2}{24
\beta_0}\frac{R_N^4}{l^4}~ \delta(\Phi_N). \label{deltarhoLyap}
\end{equation}
Here we  integrate over distances between disks instead of integration over disks positions.  The
factors $\sigma^2(\varphi_n)$ comes from double scatterings, the factor $R_{n}^2$  from
denominators of the two Green's functions, describing the propagation along $n-$th link on the
direct and return way. In writing Eq.~\eqref{deltarhoLyap} we also took into account the
renormalization of  correlated links by the $(-,-)$ pairings. As a consequence of this
renormalization $n-$th link, which is passed twice (on the direct and return paths) comes with the
factor $e^{-R_n/l}$ instead of $e^{-2R_n/l}.$ What remains to do is to express final angle $\Phi_N$
via initial angle $\Phi_0.$
 From simple geometric considerations we get
\begin{equation}
\Phi_{0}=\frac{R_1}{\sigma(\varphi_1)}\dots
\frac{R_{N}}{\sigma(\varphi_{N})}\Phi_N. \label{PhinfromPhi0}
\end{equation}
Combining Eqs.~\eqref{deltarhoLyap} and \eqref{PhinfromPhi0} and
performing the integral one gets
\begin{equation}
\frac{\delta \rho^N_{xx}}{\rho}=-\frac{1}{2}\frac{\beta^2}{\beta_0},
\qquad {\rm for} \qquad N > 1.
\end{equation}

\section{Interpretation of the resistivity correction in terms of small change of the
effective  scattering cross-section \label{interpr}}
 In our calculations of resistivity corrections  we used Eq.~\eqref{D} as a starting point. Here we briefly  discuss an
 alternative approach
 based on the accounting   of the  memory effects
 in terms of small change of the effective  cross-section. Consider first the
  $(+,+)$ ballistic returns with given values of the angles $\varphi_0$ and $\varphi_f$.
  The contribution of such  processes to the resistivity (both for $B=0$ and for $B \neq 0$)
  contains the factor ${\rm Re}~ e^{i(\varphi_0+\varphi_f)}=
 \cos(\varphi_0+\varphi_f).$ Let us introduce now the scattering angle $\varphi$ for the process $(+,+)$
 considered as a single scattering on a complex scatterer. For small values of $\varphi_0$ and
 $\varphi_f$ this angle is evidently given by $\varphi= \pi +\varphi_0+\varphi_f.$ For arbitrary
 values of $\varphi_0,\varphi_f$ the scattering angle reads
 \begin{equation}
 \varphi=\varphi_{\varphi_0,\varphi_f}= (\pi +\varphi_0+\varphi_f)({\rm mod}~2\pi ).
\label{theta}
\end{equation}
  Such definition
 ensures that $0<\varphi<2\pi.$ Next we introduce the integration over this  angle:
 $\int d\varphi_0 d\varphi_f \cos(\varphi_0+\varphi_f)\cdots = (-1)\int d\varphi \cos\varphi\int
 d\varphi_0 d\varphi_f\delta(\varphi-\varphi_{\varphi_0,\varphi_f}) \cdots  .$
 Analogous expressions can be easily written for the processes $(-,-),(+,-),$ and $(-,+).$
 As a result, one can write the contribution of four possible types   of correlations as follows
\begin{align}
&{\rm Re}\int d\varphi_0 d\varphi_f(1-e^{i\varphi_0})(1-e^{i\varphi_f}) \cdots = \nonumber
\\
 &-\int d\varphi \cos\varphi\int
 d\varphi_0 d\varphi_f\left(\delta(\varphi-\varphi_{\varphi_0,\varphi_f})+
 \delta(\varphi-\pi)
 -\delta(\varphi-\varphi_{\varphi_0,0})-\delta(\varphi-\varphi_{0,\varphi_f})\right)
 \cdots.
\label{cos}
 \end{align}
    The $(-,-)$ process is represented in Eq.~\eqref{cos} by  $\delta(\varphi-\pi).$
    Here we neglected a small  difference (on the order of $\beta_0$) between the
    angle $\varphi$ and $\pi.$
   By virtue of Eq.~\eqref{cos} one can easily check that the memory effects related to
four types of ballistic returns can be accounted quantitatively in the frame of usual Boltzmann
equation.  One should just replace the scattering cross-section on one disk $\sigma(\varphi)$ by
the effective cross-section
\begin{equation}
\sigma_{\rm eff}(\varphi)=\sigma(\varphi)+\delta\sigma_B(\varphi), \label{effective}
\end{equation}
where $\delta\sigma_B(\varphi)$ is the field-dependent correction given by
\begin{align}
\delta\sigma_B(\varphi)=&  \frac {1}{4l}  \int_a^{\infty} \frac{dR}{R}e^{-2R/l} \int_0^{2\pi} d
\varphi_0 \int_0^{2\pi} d \varphi_f \sigma(\varphi_0) \sigma(\varphi_f) e^{ nS_B }
 \nonumber
\\ &\times[\delta(\varphi-\varphi_{\varphi_0,\varphi_f})+
 \delta(\varphi-\pi)
 -\delta(\varphi-\varphi_{\varphi_0,0})-\delta(\varphi-\varphi_{0,\varphi_f})].
\label{effective1}
 \end{align}
This correction does not change the total cross-section
\begin{equation}
\int_0^{2\pi} d\varphi ~\delta\sigma_B(\varphi)=0. \label{total}
\end{equation}
In other words, the enhancement of cross-section caused by processes $(+,+)$ and $(-,-)$ is
accompanied by the reduction of the scattering due to the $(+,-)$ and $(-,+)$ correlations.  The
resistivity correction is proportional to the change of the inverse transport length
\begin{equation}
\delta\left(\frac{1}{l_{\rm tr}}\right)=n \int_0^{2\pi} d\varphi (1-\cos\varphi)
~\delta\sigma_B(\varphi)=-n \int_0^{2\pi} d\varphi \cos\varphi
~\delta\sigma_B(\varphi)\label{inverse}.
\end{equation}
 Using Eqs.~\eqref{cos}, \eqref{effective1} and \eqref{inverse} one can easily get the
Eq.~\eqref{fin2} for MR.

Finally we note that this approach is easily generalized  for calculation of the
contribution of the  trajectories having
long Lyapunov region. For such trajectories one should replace the
angles $\varphi_0$ and $\varphi_f$ in Eq.~\ref{cos} by
$\varphi_1$ and $\varphi_1' \approx 2\pi-\varphi_1$ (see Section~\ref{lyap}).
 For very  small $\beta$ such that $ \beta_0^{N+1}< \beta < \beta_0^N$ and $N \gg 1,$
 the magnetic field induced correction to the scattering cross-section is
 expressed  in a simple form
\begin{equation}
\frac{\delta \sigma_B(\varphi)-\delta \sigma_0(\varphi)}{2a} \approx
-\frac{2\beta^2\ln\beta }{9\beta_0\ln\beta_0}
\left(\delta(\varphi-\pi)-\frac{\cos^2(\varphi/2)}{\pi} \right).
\label{lyap-cross}
\end{equation}

 \section{Discussion}
\begin{figure}
\includegraphics[width=0.6\textwidth]{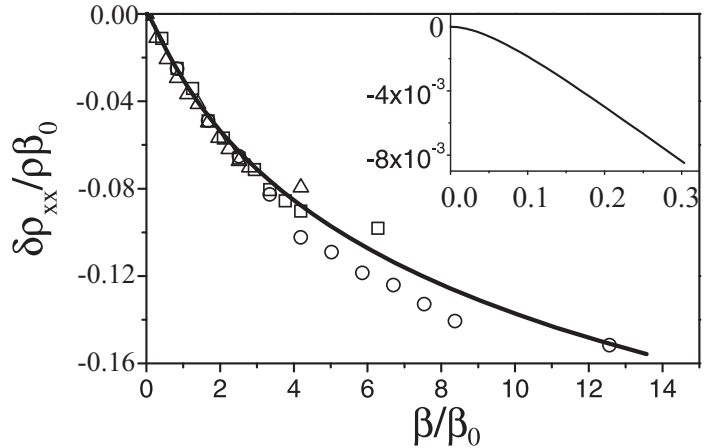}
\caption{The value of $\delta\rho_{xx}/ \rho \beta_0 $ from
Eqs.~\eqref{1},~\eqref{f(z)} (solid line) shown as a function of
$\beta/\beta_0$ together with the results of numerical simulations
\cite{dmit1} presented for different values of $\beta_0$
(triangles for $\beta_0= 0.09$ , boxes for $\beta_0= 0.06$ ,
circles for $\beta_0=0.03$). Data for all numerical curves are
shown for $\beta < 0.3.$ Inset: The crossover from quadratic to a
linear dependence at $\beta/\beta_0 \sim 0.05.$ This crossover was
not resolved in numerical simulations. } \label{fig16}
\end{figure}

In the previous sections we derived analytical theory of the
low-field anomaly in the magnetoresistance, caused by sharp
dependence of the memory effects specific for backscattering
events on the magnetic field.  Next we compare our calculations
with the results of simulations and with experiment. Note first
that there is a parametrically small nonanomalous correction to
Eq.~\eqref{1} due to returns after multiple scatterings. This
correction is given by \cite{dmit1}
$$\delta \rho_{xx}^{\prime}/\rho \approx - 0.2 \beta_0 \beta^2.$$
To compare the results of simulations \cite{dmit1} with the
theoretical results in a wider region of parameters $\beta,
\beta_0,$ we substract $\delta \rho_{xx}^{\prime}/\rho$ from the
numerical curves. Theoretical and numerical \cite{dmit1} results
are plotted in Fig.~\eqref{fig16}. in the universal units, $\delta
\rho_{xx}/ \rho \beta_0  $ versus $z=\beta/\beta_0.$ It is seen,
that the theoretical and numerical results are in a very good
agreement.
Experimental measurement of the MR in the system of antidot arrays \cite{dot1} agrees qualitatevely
with our predictions. The experimentally observed MR was linear in  magnetic field in a wide region
of fields. This region corresponds to the interval of magnetic fields, where our results can be
aproximated by linear dependence (see Eq.~\eqref{lin}). The magnetic fields used in experiment were
relatively strong, and the parabolic  asymptotic was not achieved. The quantitative comparison with
the experiment \cite{dot1} is more difficult due to several reasons. First of all, in the
structures, used in experiment, besides antidots there were also short range scatterers. Our
preliminary estimates show that accounting for a short range disorder can change the results of the
calculations. Secondly, the antidot distribution was not fully chaotic in the experiment. To be
more specific,
 the antidots
were randomly moved from the regular square  lattice distribution by
shifts
on the order of 30-40~\% of the lattice constant.
 Finally, the
antidots did not have equal sizes, the uncertainty of the size being
on the order of 50~\%.
In spite of this, a very good quantitative agreement with the experiment can be achieved by
appropriate choice of the antidot size $a$ in the uncertainty interval (see also discussion of the
experiment  in Ref.~\onlinecite{dmit1}). Note also that generalization of the theory for the case
when antidot sizes vary is straightforward. What has to be done is to average pairings $\langle
\delta \hat T^{\mu} \delta \hat T^{\nu}\rangle $  with the distribution function for the antidots
sizes. This leads to the following modification of the obtained formulae: one should change
$\sigma(\varphi_0)\sigma(\varphi_f)\to \overline{\sigma(\varphi_0)\sigma(\varphi_f)}$ in
Eq.~\eqref{fin1} and $h_B \to \overline{h_B}$ in Eq.~\eqref{hb}. Here over  line means averaging
over antidot sizes. We do not present here corresponding calculations, since the distribution
function of antidot sizes is not known for experiment. \cite{dot1}

Next we briefly discuss several interesting unsolved problems.
 Note first that above consideration of Lyapunov trajectories
 is  valid, provided that $N<1/\beta_0$.
Indeed, while calculating the $N-$th order diagrams we neglected
small contributions of $(+,-)$ and $(-,+)$ pairings on the all
correlated links. Accounting of such pairings is rather tricky and
is out of scope of this paper. We expect, however, that such
correlations may lead to a factor on the order of $(1-c\beta_0)^N$
(~$c\sim 1$) in Eq.~\eqref{PhinfromPhi0}. Therefore, for
$N>1/\beta_0$ the contribution of Lyapunov diagrams might become
$N$ dependent. This gives a low-field limit for our theory
\begin{equation}
\beta > \beta_0^{1/\beta_0}. \label{limit}
\end{equation}
We conclude that the behavior of the MR in the limit $\beta \to 0$
remains so far unclear.

In our calculations  we fully neglected quantum effects. Such
effects should decrease the "effect of empty corridor" due to the
diffraction on the edges of the disks. In the situations where the magnetic
field is not very small, $\beta > \beta_0^2$, the neglecting of
the diffraction effects
 is justified if $a \gg \sqrt{\lambda_F l}$
($\lambda_F$ is a Fermi wavelength). This criterion ensures that diffraction effects on the edges
of the disks are not relevant at the scales of the order of $l.$ In the opposite case, $a \ll
\sqrt{\lambda_F l},$ the diffraction should destroy the ``corridor effect'', thus suppressing the
anomalous MR. For small magnetic fields, when $\beta \ll \beta_0^2,$ the criterion for
negligibility of the diffraction becomes stronger, $a \gg \sqrt{\lambda_F L},$ where $L\sim l
\ln(1/\beta)/\ln(1/\beta_0) $ is the characteristic size of the Lyapunov region. Another quantum
effect which can be especially important from the point of view of the possible experimental
realizations is the weak localization phenomena.  The weak localization correction to the
conductivity also has an anomalous dependence on the magnetic field. Moreover,
 the interpretation of the memory effects in terms of small change of effective cross-section
 discussed in Section~\ref{interpr}
 is very close to  the interpretation of weak localization phenomenon developed in
 Ref.~\onlinecite{nonbk}. Similar to discussion in Section~\ref{interpr}, the coherent enhancement of
 backscattering amplitude caused by weak localization, is accompanied by reduction of coherent
 scattering in other directions, the total cross-section do not change. The competition of the non-Markovian
 and the weak localization effects might result  in new interesting phenomena. The study of
 such competition in a system with spin-orbit interaction  can be
 especially interesting, because, in contrast to the weak localization correction, the ``corridor effect'' is not
  very sensitive to spin-orbit coupling.   The detailed
analysis of the quantum effects is a challenging problem which will be addressed elsewhere.

We did not investigate the temperature dependence of the
phenomenon. This dependence is related to
the scattering by phonons (or electron-electron scattering) neglected in our
calculations.
  It worth noting that importance of the electron-phonon scattering is
  expected to
  increase with
  decreasing  magnetic field. Indeed, the potential of the electron-phonon
  interaction  depends on time,
    therefore restricting the maximal length of a trajectory with
  the Lyapunov region   $L<L_{ph},$ where $L_{ph} \sim v_F \tau_{ph} $ and
  $\tau_{ph}$ is the temperature dependent characteristic time of the electron-phonon
  scattering. This implies that at small magnetic fields one
  should replace  the logarithmic factor $\ln(\beta)/\ln(\beta_0)$,
  entering  Eq.~\eqref{ln} by a temperature dependent coefficient
  of the order of $L_{ph}/l.$
\section{Summary}
We propose a theory of the  negative
anomalous MR  in a system with strong  scatterers.   It is shown that the anomaly in the  MR arises
due to suppression of "empty corridor effect"
by  magnetic field. A detailed description of
 different types of non-Markovian
correlations related to ballistic returns    is presented.
A method of diagrammatic expansion of the
Liouville equation is developed which allows us to
 describe analytically  the effects of "empty corridor" on ballistic returns.
The analytical
expressions for anomalous MR  in different intervals of magnetic fields
 are
 derived  [see Eqs.~\eqref{1},~\eqref{f(z)},~\eqref{asimpt},~\eqref{LyapQual}]. The MR at
 very low magnetic fields  was shown to be determined by the contribution
 of electron trajectories having long Lyapunov region. An
 interpretation of the memory effects in terms of small change of
 the effective scattering cross-section is discussed. The
 analytical results are shown to be in very good agreement with
 the numerical simulations and experiment.

\section{Acknolegments}
We thank M.I.~Dyakonov for insightful discussions and R.~Jullien
for providing us with the numerical data. We are also grateful to
I.V.~Gornyi and D.G.~Polyakov for useful comments. The work was
partially supported by RFBR, grant of Russian Scientific School,
and by programmes of the RAS.

\appendix \section{\label{ApA}}
This Appendix contains the explicit expressions for the pairings $\langle \delta T^{\mu}\delta
T^{\nu}\rangle.$ Using Eqs.~\eqref{+,-}, \eqref{tt} and \eqref{Tfluke}, kernels of operators
$\delta \hat T^+ ,~ \delta  \hat T^-,$ entering in Eq.~\eqref{M1} can be written as
\begin{align}\label{kernels}
    &\delta T^+_{\alpha,\alpha^{\prime}}=
    \sigma(\alpha-\alpha^{\prime})(n^+ - n)
    \\
    \nonumber
    &\delta T^-_{\alpha,\alpha^{\prime}}=
    -\delta(\alpha-\alpha^{\prime})\delta T^-_{\alpha}
\end{align}
where
\begin{equation}
    \delta T^-_{\alpha}=\int d\alpha''
    \sigma(\alpha-\alpha'')(n^- - n)
    \label{dT-a}
\end{equation}
The functions $n^+$ and $n^-$ in Eq.~\eqref{kernels} depend on vectors $\mathbf a(\mathbf n',
\mathbf n )$ and $\mathbf a(\mathbf n'', \mathbf n )$ correspondingly (see Eqs.~\eqref{npm},
\eqref{adef}).
The pairings of operators  $ \delta \hat T^{\mu}$ and $ \delta \hat T^{\nu}$ can be calculated with
the use of Eqs.~\eqref{pair},\eqref{kernels},\eqref{dT-a}. As a result we have
\begin{align} \label{4++}
\langle \delta &T^+_{\beta,\alpha_2}(\mathbf r_2) \delta
T^+_{\alpha_1, \beta_0}(\mathbf r_1)\rangle =
\sigma(\beta-\alpha_2)\sigma(\alpha_1-\beta_0)
J(-\mathbf a_{\beta,\alpha_2},-\mathbf
a_{\alpha_1,\beta_0})
\end{align}
\begin{align} \label{4--}
\langle \delta &T^-_{\beta}(\mathbf r_2) \delta T^-_{
\beta_0}(\mathbf r_1)\rangle
=
\int d \alpha' d \alpha''
\sigma(\beta-\alpha')\sigma(\beta_0-\alpha'')
J(\mathbf a_{\beta,\alpha'},\mathbf a_{\beta_0,\alpha''})
\end{align}
\begin{align} \label{4+-}
\langle \delta &T^+_{\beta,\alpha_2}(\mathbf r_2) \delta T^-_{
\beta_0}(\mathbf r_1)\rangle = \sigma(\beta-\alpha_2)
\int d\alpha'\sigma(\beta_0-\alpha')J(-\mathbf
a_{\beta,\alpha_2},\mathbf a_{\beta_0,\alpha'})
\end{align}
\begin{align}\label{4-+}
\langle \delta &T^-_{\beta}(\mathbf r_2) \delta T^+_{\alpha_1,
\beta_0}(\mathbf r_1)\rangle =
 \sigma(\alpha_1-\beta_0)
\int d\alpha' \sigma(\beta - \alpha') J(\mathbf a_{\beta,\alpha'},-\mathbf
a_{\alpha_1,\beta_0}).
\end{align}
 Here function $J$ emerges as a result of pairings of $\langle (n^{\mu}-n)
 (n^{\nu} - n)\rangle$ (see Eq.~\eqref{pair}) and is given by
\begin{equation}
J(\mathbf a, \mathbf b)= n\int d\mathbf R \delta(\mathbf
r_2-\mathbf R - \mathbf a) \delta(\mathbf r_1-\mathbf R - \mathbf
b) \label{J}
\end{equation}

\section{\label{ApB}}
To calculate $ \left \langle \delta T^-_{\alpha[\mathbf r_2]}(\mathbf r) \delta T^-_{
\alpha[-\mathbf r_1]} (\mathbf r^{\prime})\right \rangle$  we first do the integral in the right
hand side of Eq.~\eqref{dT-a}. Integral of the term containing average density $n$ is trivial and
reduces to $1/l.$ In the integral containing $n^-$ it is convenient to introduce orthogonal
coordinates $x$ and $y$ such that $x$ axis is collinear with the velocity of the incident particle
and write the delta function entering the definition \eqref{npm} of $n^-$   as $\delta(\mathbf
r-\mathbf R_i-\mathbf a)=\delta ( y- R_i^y- a_y) \delta ( x- R_i^x- a_x).$ We proceed by lifting
the $y$ dependent delta function by integration over the angle $\alpha'.$ It is clear that the
integral is vanishing for $\vert y-R_i^y \vert >a.$ To calculate the integral for smaller $a$ note
that by virtue of Eq.~\eqref{adef} $ \vert {d a_y}/{d \alpha'}\vert= \vert d( \mathbf a \times
\mathbf n)/d \alpha' \vert=\sigma(\alpha-\alpha') $ and therefore the integral of the $y$ dependent
delta function cancels the scattering cross-section in Eq.~\eqref{dT-a}. As a result we get
\begin{equation}
\delta T^-_{\alpha}(\mathbf r) = 1/l -\sum_i\delta (x-R_i^x- a_x)
\theta_i \approx1/l -\sum_i \delta (x-R_i^x) \theta_i
\label{deltaT-}
\end{equation}
Here $\theta_i$ is a unit step function which is equal to unity when $\vert y-R_i^y \vert <a$ and
vanishes otherwise. Using Eq.~\eqref{deltaT-} we can write for backscattering angles close to $\pi$
($\Phi \sim a/l$)
\begin{equation}
\left \langle \delta T^-_{\alpha[\mathbf r_2 ]}(\mathbf r) \delta
T^-_{ \alpha[-\mathbf r_1]} (\mathbf r^{\prime})\right \rangle
\approx n h(x,\Phi)\delta(x-x^{\prime})\approx n
h(r,\Phi)\delta(r-r^{\prime}), \label{dTmap1}
\end{equation}
where $$h(r, \Phi) = \left(2a - \left\vert \Phi   \right\vert
r\right) \theta \left(2a - \left\vert \Phi  \right\vert r\right)$$
 is shown in Fig.~\ref{fig12}.


\begin{thebibliography}{99}
\bibitem{off} For $\beta \ll 1$ one can neglect the
off-diagonal part of the conductivity tensor.
\end{thebibliography}
\end{document}